\documentclass[a4paper,pra,showpacs,
nofootinbib,reprint]{revtex4-1}
\usepackage[utf8]{inputenc}
\usepackage[T1]{fontenc}
\usepackage[UKenglish]{babel}
\usepackage{lmodern}
\usepackage[colorlinks]{hyperref}
\usepackage{graphicx}
\usepackage{amsmath,amssymb,amsthm}
\usepackage{xspace}
\usepackage{mathtools}
\usepackage{dsfont}
\usepackage{lipsum}
             \usepackage{siunitx}
             \usepackage{pgf,tikz}
             \usepackage{pgfplots}

\usepackage{paralist}

\usepackage{enumitem}
\usepackage[normalem]{ulem}
\usepackage{units}

\usepackage{tabularx}

\DeclareMathOperator{\sgn}{sgn}

\DeclareMathOperator{\tr}{tr}

\DeclareMathOperator{\conv}{conv}

\let\originalleft\left
\let\originalright\right
\renewcommand{\left}{\mathopen{}\mathclose\bgroup\originalleft}
\renewcommand{\right}{\aftergroup\egroup\originalright}

\newcommand{\KetBra}[1]{{\Ket{#1}\!\Bra{#1} }}

\newcommand{\Bra}[1]{{ \langle \! \langle{#1}\vert }}
\newcommand{\Ket}[1]{{ \vert {#1}  \rangle \!  \rangle}}

\newcommand{\bra}[1]{\left\langle #1 \right|}
\newcommand{\ket}[1]{\left| #1 \right\rangle}

\newcommand{\ketbra}[2]{\left|#1\middle\rangle\!\middle\langle#2\right|}
\newcommand{\proj}[1]{\left|#1\middle\rangle\!\middle\langle#1\right|}

\newcommand{\wsepcone}{{\mathcal W}^{\text{sep}}}

\newcommand{\mathand}{\quad\text{and}\quad}

\newcommand{\Hi}{\mathcal{H}}

\newcommand{\id}{\mathds{1}}

\renewcommand{\ketbra}[2]{\left|#1\middle\rangle\!\middle\langle#2\right|}
\renewcommand{\proj}[1]{\left|#1\middle\rangle\!\middle\langle#1\right|}

\definecolor{mycolor1}{rgb}{0.2472, 0.24, 0.6}
\definecolor{mycolor2}{rgb}{0.6, 0.24, 0.4429}

\hypersetup{pdftitle={{Witnesses of causal nonseparability: an introduction and a few case studies}}}

\begin{document}

\title{Witnesses of causal nonseparability: \\ an introduction and a few case studies}

\author{Cyril Branciard}
\affiliation{Institut N\'eel, CNRS and Universit\'e Grenoble Alpes, 38042 Grenoble Cedex 9, France}

\date{\today}

\begin{abstract}

It was recently realised that quantum theory allows for so-called \emph{causally nonseparable processes}, which are incompatible with any definite causal order. 
This was first suggested on a rather abstract level by the formalism of \emph{process matrices}, which only assumes that quantum theory holds locally in some observers' laboratories, but does not impose a global causal structure; it was then shown, on a more practical level, that the \emph{quantum switch}---a new resource for quantum computation that goes beyond causally ordered circuits---provided precisely a physical example of a causally nonseparable process.
To demonstrate that a given process is causally nonseparable, we introduced in [Ara\'ujo \emph{et al.}, \href{http://dx.doi.org/10.1088/1367-2630/17/10/102001}{{\it New J. Phys.} {\bf 17}, 102001 (2015)}] the concept of \emph{witnesses of causal nonseparability}.
Here we present a shorter introduction to this concept, and concentrate on some explicit examples to show how to construct and use such witnesses in practice.

\end{abstract}

\maketitle

\section{Introduction}

In our common understanding of the world, we typically perceive events as happening one after another, in a given order. Relations between events are understood in terms of causes and effects, where a cause can only precede an effect. Events can thus be embedded in a \emph{causal structure}, which defines the \emph{causal order} between them.

This viewpoint is ingrained for instance in the circuit model for computation or information processing, where operations are performed by gates that are applied in a definite order. While the assumption that events follow a definite causal order seems natural in the classical world, one may nevertheless wonder whether it must really always be so. One may in particular become suspicious when entering the quantum world, where the properties of physical systems are not always well-defined.

A general framework, that of \emph{process matrices}, was recently introduced to investigate physical processes without pre-assuming a definite global causal structure; the framework only assumes that quantum theory correctly describes what happens locally, in some observers' laboratories~\cite{oreshkov12}. It was shown that this allows for processes that are incompatible with any definite causal order---so-called \emph{causally nonseparable processes}. The framework was first introduced on a rather abstract level, with no clear physical interpretation given to the first examples of causally nonseparable processes.
However, a concrete physical example of a causally nonseparable process was later exhibited~\cite{araujo15,oreshkov15}: namely, the recently proposed \emph{quantum switch}, a new resource for quantum computation where the order of operations is controlled by a qubit in a superposition of two different states---which indeed does not fit in the standard framework of causally ordered quantum circuits~\cite{chiribella09}.

To ensure that this notion of causal nonseparability has any practical meaning, one needs of course to be able to verify that a given process is causally nonseparable. This was first done in Ref.~\cite{oreshkov12} through the violation of a \emph{causal inequality}---an inequality bounding the correlations compatible with a definite causal order, and whose violation can only be obtained from a causally nonseparable process. This is however a very strong argument for causal nonseparability. In fact, not all causally nonseparable processes violate a causal inequality; the quantum switch indeed provides such an example~\cite{araujo15,oreshkov15}.

More recently we introduced, in analogy with entanglement witnesses, the concept of \emph{witnesses of causal nonseparability} (or \emph{causal witnesses}, as we initially called them)~\cite{araujo15}. Here a witness corresponds to an operator that can (in principle) be `measured' on a given process by combining the statistics of various operations, and whose expectation value, if negative, certifies the causal nonseparability of the process. We showed in particular that a witness can be efficiently constructed for any causally nonseparable process.

The objective of this paper is to present a somewhat shorter introduction to this new concept of witnesses of causal nonseparability. We will avoid here some of the technicalities in the proofs, and refer directly to Ref.~\cite{araujo15} for that. We will then present several different explicit examples of causally nonseparable processes and of witnesses---in particular for the quantum switch, investigating its robustness to different kinds of noise---so as to illustrate how to construct and use them in practice.

\section{The process matrix formalism}

\subsection{In the general bipartite case}

Consider an experiment with two parties, Alice and Bob, sitting in closed laboratories and exchanging physical systems. In a single run of the experiment, each party opens their lab only once to let some incoming system enter, and once to send some outgoing system out. They can perform some operation on these systems, which may output some result $a$ for Alice and $b$ for Bob.

While we do not pre-suppose a definite causal order between the events happening in Alice and Bob's labs, we assume that what happens \emph{locally} inside the labs is correctly described by quantum theory. That means, we can attach some Hilbert spaces $\mathcal H^{A_I}$ and $\mathcal H^{B_I}$ to their incoming systems and some Hilbert spaces $\mathcal H^{A_O}$ and $\mathcal H^{B_O}$ to their outgoing systems, and their choices of operations correspond to so-called quantum instruments~\cite{davies70}---i.e., sets of completely positive (CP) maps which sum up to CP and trace-preserving maps~\cite{chuang00}. These can conveniently be represented, using the Choi-Jamio{\l}kowski (CJ) isomorphism, by positive semidefinite matrices $M_{a}^{A_IA_O} \in A_I\otimes A_O$ and $M_{b}^{B_IB_O} \in B_I\otimes B_O$, where $A_I$ and $A_O$ (resp. $B_I$ and $B_O$) denote the spaces of Hermitian linear operators over Alice's (Bob's) incoming and outgoing Hilbert spaces, and where the subscripts refer to the outcomes $a,b$ they correspond to.
To define valid instruments, these matrices must satisfy
\begin{equation}
\begin{gathered}
 M^{A_I A_O}_{a}\geq 0 \, , \quad
 \tr_{A_O} \big[{\textstyle \sum_{a}} M^{A_I A_O}_{a}\big] = \id^{A_I} \, , \\
 M^{B_I B_O}_{b}\geq 0 \, , \quad
 \tr_{B_O} \big[{\textstyle \sum_{b}} M^{B_I B_O}_{b}\big] = \id^{B_I} \, ,
 \label{instrument}
\end{gathered}
\end{equation}
where $\id^X$ denotes the identity operator in the space $X$ (in general, superscripts on operators will refer to the space they are acting on) and $\tr_X$ is the partial trace over $X$.
In this paper we will only consider finite-dimensional Hilbert spaces; the dimension of a Hilbert space $\mathcal H^X$ will be denoted $d_X$.

\subsubsection{Process matrices}

\begin{figure}
  \centering
  \begin{tikzpicture}[scale=1.6]
		\node[draw, thick, rectangle,minimum width=0.7cm,minimum height=0.9cm, fill=white!80!gray] (S) at (-1,0) {\footnotesize $M_a^{A_I\!A_O}$};
		\node[red,thick, minimum width=0.7cm,minimum height=0.9cm] (W) at (0,0) {$W$};
		\node[draw, thick, rectangle,minimum width=0.7cm,minimum height=0.9cm, fill=white!80!gray] (T) at (1,0) {\footnotesize $M_b^{B_I\!B_O}$};
                \draw[red, thick] (-1.2,-1) -- (1.2,-1) -- (1.2,-0.6) -- (0.35,-0.6) -- (0.35,0.6) -- (1.2,0.6) -- (1.2,1) -- (-1.2,1) -- (-1.2, 0.6) -- (-0.35,0.6) -- (-0.35,-0.6) -- (-1.2,-0.6) -- (-1.2,-1);
                \draw[->, red, thick] (S) -- (-1,0.6);
                \draw[<-, red, thick] (S) -- (-1,-0.6);
                \draw[->, red, thick] (T) -- (1,0.6);
                \draw[<-, red, thick] (T) -- (1,-0.6);
		\node[red] (SO) at ([shift={(0.27cm,0.45cm)}]S) {\footnotesize $\Hi^{A_O}$};
		\node[red] (SI) at ([shift={(0.27cm,-0.45cm)}]S) {\footnotesize $\Hi^{A_I}$};
		\node[red] (TO) at ([shift={(-0.25cm,0.45cm)}]T) {\footnotesize $\Hi^{B_O}$};
		\node[red] (TI) at ([shift={(-0.25cm,-0.45cm)}]T) {\footnotesize $\Hi^{B_I}$};
		\node[black, rotate=90] (A) at ([shift={(-0.52cm,0cm)}]S) {Alice};
		\node[black, rotate=270] (B) at ([shift={(+0.52cm,0cm)}]T) {Bob};
  \end{tikzpicture}
  \caption{Two parties, Alice and Bob, perform some quantum operations $M_a^{A_IA_O}$ and $M_b^{B_IB_O}$---some CP maps with outcomes $a$, $b$---which act on some incoming systems in the Hilbert spaces $\Hi^{A_I}$, $\Hi^{B_I}$ and generate some outgoing systems in the Hilbert spaces $\Hi^{A_O}$, $\Hi^{B_O}$. The \emph{process matrix} $W$ represents the physical resource that connects their labs, generalising the notions of quantum states and of quantum channels.}
  \label{fig:W}
\end{figure}
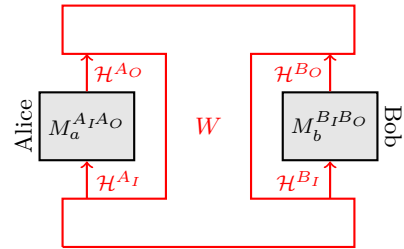

The correlations established by Alice and Bob in such a scenario can be described by the probabilities $P( M_{a}^{A_IA_O}, M_{b}^{B_IB_O})$ that Alice and Bob obtain the outcomes $a,b$ attached to the CP maps $M_{a}^{A_IA_O}, M_{b}^{B_IB_O}$. 
As shown in~\cite{oreshkov12}, these correlations can be written in the form
\begin{eqnarray}
P\big(M_{a}^{A_IA_O}\!, M_{b}^{B_IB_O}\big) = \tr \left[ \big(M_{a}^{A_IA_O}\otimes M_{b}^{B_IB_O} \big) \cdot W \right] \qquad
\label{born}
\end{eqnarray}
(with $\tr$ now denoting the full trace), for some Hermitian matrix $W \in A_I\otimes A_O \otimes B_I\otimes B_O$. This so-called \emph{process matrix} is the central object of the formalism; it describes the physical resource (the \emph{process}) that connects Alice and Bob's labs, and generalises both the notion of a quantum state---in which case Eq.~\eqref{born} reduces to the standard Born rule---and of a quantum channel; see Figure~\ref{fig:W}.

Not all matrices $W\in A_I\otimes A_O \otimes B_I\otimes B_O$ define valid processes. As one can show~\cite{oreshkov12,araujo15}, the constraint that all probabilities obtained through~\eqref{born} must be nonnegative and normalised (including in situations where Alice and Bob may share and interact with ancillary entangled systems) leads to the following conditions that valid process matrices must satisfy:
\\
\noindent
\begin{subequations}
\begin{minipage}{0.5\linewidth}
\begin{align}
{}_{[1-B_O]A_IA_O}W &= 0 \, , \label{eq:valid_cond_bi_1BAA} \ \\
{}_{[1-A_O]B_IB_O}W &= 0 \, , \label{eq:valid_cond_bi_1ABB} \ \\
{}_{[1-A_O][1-B_O]}W &= 0 \, , \label{eq:valid_cond_bi_1A1B} \ \\[-2mm] \nonumber
\end{align}
\end{minipage}
\quad
\begin{minipage}{0.4\linewidth}
\begin{align}
W &\ge 0 \, , \label{eq:valid_cond_bi_pos} \\
\tr W &= d_O \, , \label{eq:valid_cond_bi_norm} \\[-2mm] \nonumber
\end{align}
\end{minipage}
\end{subequations}
\\
\noindent with $d_O = d_{A_O}d_{B_O}$ and where we used (and will use throughout the paper) the following notation, introduced in~\cite{araujo15}:
\begin{equation}
\begin{gathered}
 _X W = \frac{\id^{X}}{d_X} \otimes \tr_X W \, , \quad {}_1 W = W \, , \\
 {}_{\big[\sum_i \alpha_i X_i\big]} W = \sum_i \, \alpha_i \ {}_{X_i\!} W \, . \quad
\end{gathered}
\end{equation}

Eqs.~\eqref{eq:valid_cond_bi_1BAA}--\eqref{eq:valid_cond_bi_1A1B} define a linear subspace $\mathcal L_V \subset A_I\otimes A_O \otimes B_I\otimes B_O$, which valid process matrices belong to. Eq.~\eqref{eq:valid_cond_bi_pos} tells us that process matrices are in the set $\mathcal P$ of positive semidefinite matrices.
We shall often ignore, for convenience, the normalisation condition~\eqref{eq:valid_cond_bi_norm}, and define the set of nonnormalised process matrices as $\mathcal W = \mathcal L_V \cap \mathcal P$; as can easily be checked, this set is a closed convex cone.

\subsubsection{Causally separable vs causally nonseparable processes}

Processes that do not allow Bob to signal to Alice are compatible with a causal order where Alice acts before Bob, which we write $A \prec B$. We shall generically denote by $W^{A \prec B}$ the corresponding process matrices; these simply represent standard, causally ordered quantum circuits. One can show that these are the matrices in $A_I\otimes A_O \otimes B_I\otimes B_O$, which satisfy~\cite{gutoski06,chiribella09b,araujo15}
\\
\noindent
\begin{subequations}
\begin{minipage}{0.55\linewidth}
\begin{align}
{}_{[1-B_O]}W^{A \prec B} &= 0 \, , \label{eq:valid_cond_biAB_1B} \\
{}_{[1-A_O]B_IB_O}W^{A \prec B} &= 0 \, , \label{eq:valid_cond_biAB_1ABB} \\[-1mm] \nonumber
\end{align}
\end{minipage}
\ 
\begin{minipage}{0.42\linewidth}
\begin{align}
W^{A \prec B} &\ge 0 \, , \label{eq:valid_cond_biAB_pos} \\
\tr W^{A \prec B} &= d_O \, . \ \label{eq:valid_cond_biAB_norm} \\[-1mm] \nonumber
\end{align}
\end{minipage}
\end{subequations}
\\
\noindent Note that Eqs.~\eqref{eq:valid_cond_biAB_1B}--\eqref{eq:valid_cond_biAB_1ABB} imply Eqs.~\eqref{eq:valid_cond_bi_1BAA}--\eqref{eq:valid_cond_bi_1A1B}, which ensures that the $W^{A \prec B}$ matrices thus characterised are valid process matrices.
Eqs.~\eqref{eq:valid_cond_biAB_1B}--\eqref{eq:valid_cond_biAB_1ABB} thus define a linear subspace $\mathcal L_{A \prec B} \subset \mathcal L_V$. Together with Eq.~\eqref{eq:valid_cond_biAB_pos}, we can define the closed convex cone of nonnormalised process matrices compatible with the causal order $A \prec B$, as ${\cal W}^{A \prec B} = \mathcal L_{A \prec B} \cap \mathcal P$.

Similarly, processes that do not allow Alice to signal to Bob are compatible with a causal order $B \prec A$, where Bob acts before Alice. The corresponding process matrices $W^{B \prec A}$ (which again simply represent standard, causally ordered quantum circuits) satisfy
\\
\noindent
\begin{subequations}
\begin{minipage}{0.55\linewidth}
\begin{align}
{}_{[1-A_O]}W^{B \prec A} &= 0 \, , \label{eq:valid_cond_biBA_1A} \\
{}_{[1-B_O]A_IA_O}W^{B \prec A} &= 0 \, , \label{eq:valid_cond_biBA_1BAA} \\[-1mm] \nonumber
\end{align}
\end{minipage}
\ 
\begin{minipage}{0.42\linewidth}
\begin{align}
W^{B \prec A} &\ge 0 \, , \label{eq:valid_cond_biBA_pos} \\
\tr W^{B \prec A} &= d_O \, . \ \label{eq:valid_cond_biBA_norm} \\[-1mm] \nonumber
\end{align}
\end{minipage}
\end{subequations}
\\
\noindent Eqs.~\eqref{eq:valid_cond_biBA_1A}--\eqref{eq:valid_cond_biBA_1BAA} define a linear subspace $\mathcal L_{B \prec A} \subset \mathcal L_V$. Together with Eq.~\eqref{eq:valid_cond_biBA_pos}, we define the closed convex cone of nonnormalised process matrices compatible with the causal order $B \prec A$, as ${\cal W}^{B \prec A} = \mathcal L_{B \prec A} \cap \mathcal P$.

\medskip

One can still easily make sense of a convex mixture
\begin{equation}
 W^{\text{sep}} \, = \, q \, W^{A \prec B} \, + \, (1{-}q) \, W^{B \prec A} \, , \label{def:caus_sep_bi}
\end{equation}
representing a process that is compatible with the causal order $A \prec B$ with some probability $q \in [0,1]$, and compatible with the causal order $A \prec B$ with some probability $1-q$. Process matrices that can be decomposed in this form (or directly, the process they represent) are said to be \emph{causally separable}.
Ignoring again the normalisation constraint, the set of nonnormalised causally separable process matrices also forms a closed convex cone, obtained as the Minkowski sum\footnote{In Ref.~\cite{araujo15} we wrote, equivalently, $\wsepcone = \conv( {\cal W}^{A \prec B} \cup {\cal W}^{B \prec A} )$, where $\conv$ denotes the convex hull.}
\begin{eqnarray}
\wsepcone &=& {\cal W}^{A \prec B} + {\cal W}^{B \prec A} \nonumber \\
&=& \Big\{W^{A \prec B} + W^{B \prec A} \Big| W^{A \prec B} \in {\cal W}^{A \prec B} , \nonumber \\[-2mm]
&& \hspace{3cm} W^{B \prec A} \in {\cal W}^{B \prec A}\Big\} \, . \qquad \label{eq:Wsep_cone_bi}
\end{eqnarray}

As first proven in~\cite{oreshkov12}, there exist valid process matrices that \emph{cannot} be decomposed as in~\eqref{def:caus_sep_bi}, and which are therefore not in $\wsepcone$. These are called \emph{causally nonseparable}, and represent processes that are incompatible with any definite causal order---be it well-defined, or only determined with some probability.

\subsection{In a particular tripartite scenario}

The scenario considered before can be generalised to more parties. While it is fairly easy to construct and characterise multipartite process matrices~\cite{oreshkov12,baumeler13,araujo15}, defining the notion of causal (non)separability is somewhat more subtle in such a setting~\cite{oreshkov15,branciard16b}. In Ref.~\cite{araujo15} we restricted our study to a specific tripartite scenario, whose analysis matches that in the bipartite case quite closely (note indeed the similarities between the equations below and those in the previous subsection). We will again restrict ourselves to that case here, which is already quite relevant in practice, as we will see with the example of the quantum switch in Subsection~\ref{subsec_Wswitch}.

\subsubsection{Process matrices}

In this particular scenario, the third party we introduce, Charlie, only has an incoming system in a Hilbert space $\mathcal H^{C_I}$ (as before, we will denote by $d_{C_I}$ its dimension, and by $C_I$ the space of Hermitian linear operators acting on $\mathcal H^{C_I}$), with no outgoing system---or equivalently: Charlie has a trivial outgoing system, in a trivial Hilbert space $\mathcal H^{C_O} \equiv \mathbb{C}$ of dimension $d_{C_O} = 1$.
For a CP map $M_{c}^{C_I}$ applied by Charlie, which reduces here to an element of a positive operator-valued measure (POVM)~\cite{chuang00,araujo15}, the generalised Born rule~\eqref{born} simply becomes
\begin{eqnarray}
&& P\big(M_{a}^{A_IA_O}\!, M_{b}^{B_IB_O}\!, M_{c}^{C_I}\big) \nonumber \\
&& \qquad = \tr \left[ \big(M_{a}^{A_IA_O}\otimes M_{b}^{B_IB_O} \otimes M_{c}^{C_I} \big) \cdot W \right] \, , \qquad
\label{born3}
\end{eqnarray}
with now a process matrix $W$ in $A_I\otimes A_O \otimes B_I\otimes B_O \otimes C_I$.

Valid process matrices in this scenario satisfy~\cite{araujo15}
\\
\noindent
\begin{subequations}
\begin{minipage}{0.55\linewidth}
\begin{align}
{}_{[1-B_O]A_IA_OC_I}W &= 0 \, , \label{eq:valid_cond_tri_1BAAC} \\
{}_{[1-A_O]B_IB_OC_I}W &= 0 \, , \label{eq:valid_cond_tri_1ABBC} \\
{}_{[1-A_O][1-B_O]C_I}W &= 0 \, , \label{eq:valid_cond_tri_1A1BC} \ \\[-2mm] \nonumber
\end{align}
\end{minipage}
\quad
\begin{minipage}{0.37\linewidth}
\begin{align}
W &\ge 0 \, , \label{eq:valid_cond_tri_pos} \\
\tr W &= d_O \, , \ \label{eq:valid_cond_tri_norm} \\[-2mm] \nonumber
\end{align}
\end{minipage}
\end{subequations}
\\
\noindent with again $d_O = d_{A_O}d_{B_O}$.
Eqs.~\eqref{eq:valid_cond_tri_1BAAC}--\eqref{eq:valid_cond_tri_1A1BC} define, as before, a linear subspace $\mathcal L_V \subset A_I\otimes A_O \otimes B_I\otimes B_O \otimes C_I$. We can again characterise the closed convex cone of nonnormalised process matrices as $\mathcal W = \mathcal L_V \cap \mathcal P$.

\subsubsection{Causally separable vs causally nonseparable processes}

Since we assume that Charlie does not send any outgoing system out of his lab, one can argue~\cite{araujo15} that the only relevant causal orders are those where he is last; we are thus left to consider only the orders $A \prec B \prec C$ and $B \prec A \prec C$.

The process matrices $W^{A \prec B \prec C}$ that are compatible with the causal order $A \prec B \prec C$ (and which thus, again, simply represent standard, causally ordered quantum circuits) are those, which satisfy~\cite{gutoski06,chiribella09b,araujo15}
\begin{subequations}
\begin{align}
{}_{[1-B_O]C_I}W^{A \prec B \prec C} &= 0 \, , \label{eq:valid_cond_triABC_1BC} \\
{}_{[1-A_O]B_IB_OC_I}W^{A \prec B \prec C} &= 0 \, , \label{eq:valid_cond_triABC_1ABBC} \\[2mm]
W^{A \prec B \prec C} &\ge 0 \, , \label{eq:valid_cond_triABC_pos} \\
\tr W^{A \prec B \prec C} &= d_O \, . \label{eq:valid_cond_triABC_norm}
\end{align}
\end{subequations}
Eqs.~\eqref{eq:valid_cond_triABC_1BC}--\eqref{eq:valid_cond_triABC_1ABBC} define here a linear subspace $\mathcal L_{A \prec B \prec C} \subset \mathcal L_V$. Together with Eq.~\eqref{eq:valid_cond_triABC_pos}, we define the closed convex cone of nonnormalised process matrices compatible with the causal order $A \prec B \prec C$, as ${\cal W}^{A \prec B \prec C} = \mathcal L_{A \prec B \prec C} \cap \mathcal P$.

Similarly, the process matrices $W^{B \prec A \prec C}$ that are compatible with the causal order $B \prec A \prec C$ are those which satisfy
\begin{subequations}
\begin{align}
{}_{[1-A_O]C_I}W^{B \prec A \prec C} &= 0 \, , \label{eq:valid_cond_triBAC_1AC} \\
{}_{[1-B_O]A_IA_OC_I}W^{B \prec A \prec C} &= 0 \, , \label{eq:valid_cond_triBAC_1BAAC} \\[2mm]
W^{B \prec A \prec C} &\ge 0 \, , \label{eq:valid_cond_triBAC_pos} \\
\tr W^{B \prec A \prec C} &= d_O \, . \label{eq:valid_cond_triBAC_norm}
\end{align}
\end{subequations}
Eqs.~\eqref{eq:valid_cond_triBAC_1AC}--\eqref{eq:valid_cond_triBAC_1BAAC} define a linear subspace $\mathcal L_{B \prec A \prec C} \subset \mathcal L_V$. The closed convex cone of nonnormalised process matrices compatible with the causal order $B \prec A \prec C$ is defined here as ${\cal W}^{B \prec A \prec C} = \mathcal L_{B \prec A \prec C} \cap \mathcal P$.

\medskip

In analogy with the previous case, any process matrix in the present scenario that can be decomposed as 
\begin{equation}
 W^{\text{sep}} \, = \, q \, W^{A \prec B \prec C} \, + \, (1{-}q) \, W^{B \prec A \prec C} \, , \label{def:caus_sep_tri}
\end{equation}
with $q \in [0,1]$, is called \emph{causally separable}\footnote{This definition of causal separability was proposed in~\cite{araujo15} for the particular tripartite case we consider here. Note that it is different from that proposed in~\cite{oreshkov15} for general multipartite processes. As can actually be shown~\cite{branciard16b}, our definition here rather matches the notion of \emph{extensible causal separability} of Ref.~\cite{oreshkov15}.}.
The set of nonnormalised causally separable process matrices also forms a closed convex cone, which can again be expressed here as the Minkowski sum
\begin{eqnarray}
\wsepcone &=& {\cal W}^{A \prec B \prec C} + {\cal W}^{B \prec A \prec C} \, . \label{eq:Wsep_cone_tri}
\end{eqnarray}

Process matrices that \emph{cannot} be decomposed as in~\eqref{def:caus_sep_tri}, and are thus not in $\wsepcone$, are called \emph{causally nonseparable}. These are incompatible with any definite causal order (with Charlie last)---be it well-defined, or only determined with some probability.

\section{Witnesses of causal nonseparability}

\subsection{Definition and characterisation}

The concept of causal nonseparability represents a new type of resource compatible (at least locally) with quantum theory, which allows us to go beyond the standard framework of causally ordered quantum circuits~\cite{chiribella09}.
An important question, to ensure this concept has some concrete physical ground, is: how to detect it and verify it in practice?

One possible approach, used by Oreshkov \emph{et al.} in~\cite{oreshkov12}, is through the violation of a so-called \emph{causal inequality}---namely, a bound on the correlations that are compatible with a definite causal order. Since all correlations generated by causally separable processes must satisfy such an inequality, a violation indeed ensures that the underlying process is causally nonseparable.
Note that such a demonstration is \emph{device-independent}, in the sense that one only looks at the observed correlations, without making assumptions on what operations the devices perform. Violating a causal inequality is however quite a strong requirement. In fact, just as not all entangled quantum states violate a Bell inequality~\cite{werner89,barrett02}, not all causally nonseparable processes violate a causal inequality~\cite{araujo15,oreshkov15} (an example being the quantum switch described below): one must then use less stringent criteria to detect causal nonseparability.

\begin{figure}
 \includegraphics[width=\columnwidth]{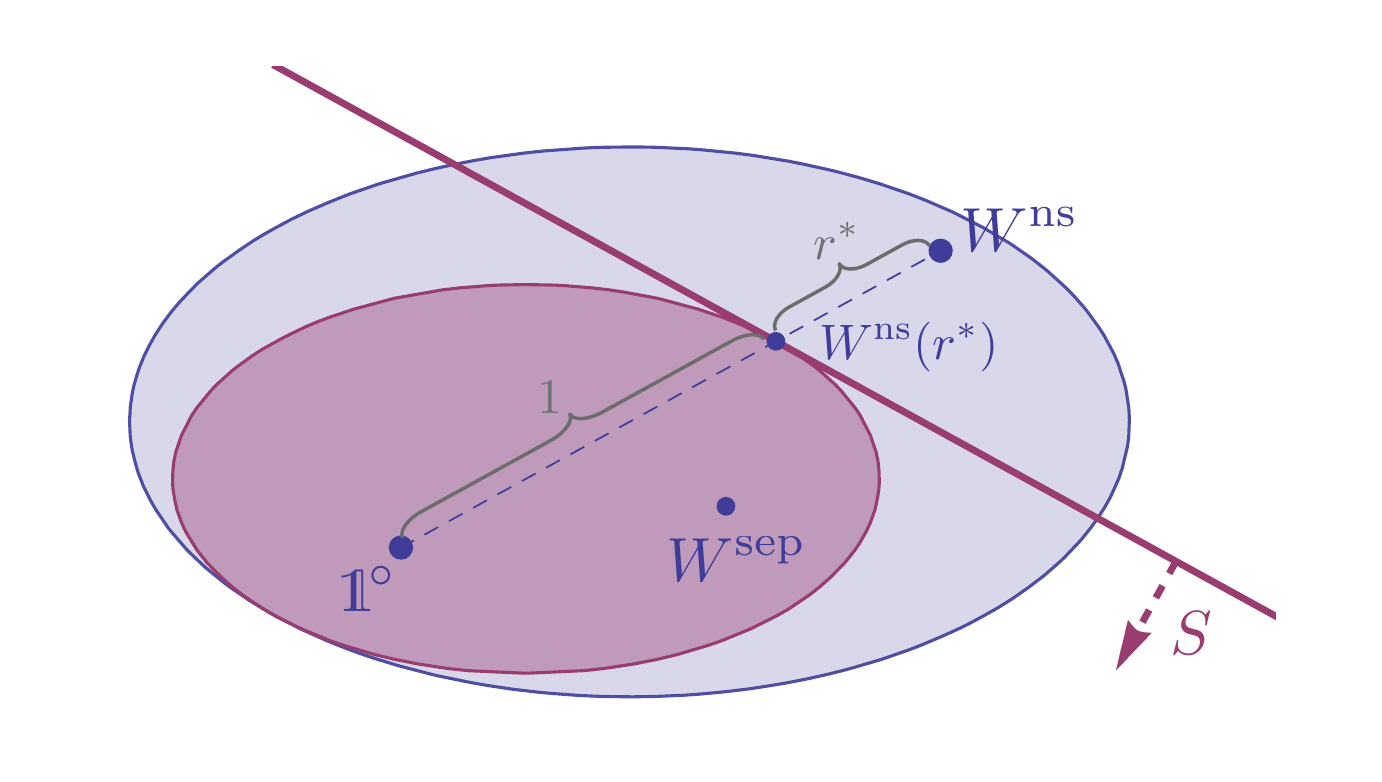}
\caption{The set of causally separable process matrices, schematically represented by the inner ellipse, is closed and convex. From the separating hyperplane theorem, for any causally nonseparable process matrix $W^{\text{ns}}$ (in the larger ellipse containing all valid process matrices), there exists a hyperplane, represented by the solid line, that separates it from all causally separable process matrices $W^{\text{sep}}$. That is, there exists a Hermitian operator $S$---a \emph{witness of causal nonseparability}---such that $\tr [S \cdot W^{\text{sep}} ] \ge 0$ for all $W^{\text{sep}}$, but $\tr [S \cdot W^{\text{ns}} ] < 0$.
\\
Solving the SDP problems of Subsection~\ref{subsec:SDPs} provides such a witness, which is optimal with respect to the resistance of $W^{\text{ns}}$ to white noise, represented by the process matrix $\id^\circ$: as depicted on the Figure, it detects the causal nonseparability of all process matrices $W^{\text{ns}}(r) = \frac{1}{1+r}( W^{\text{ns}} + r \, \id^\circ)$ for $r$ lower than the random robustness $r^*$ (directly obtained as a result of the SDP optimisation) above which $W^{\text{ns}}(r)$ becomes causally separable.
}
\label{fig:witness}
\end{figure}

In Ref.~\cite{araujo15} we introduced for that, in analogy with entanglement witnesses~\cite{horodecki96,terhal00}, the concept of \emph{witnesses of causal nonseparability}---which we simply abbreviated (somewhat abusively) to \emph{causal witnesses}.
In this context, a witness is defined as any Hermitian operator $S$ such that 
\begin{eqnarray}
\tr[S \cdot W^\text{sep}] \ge 0 \label{eq:trSWgeq0}
\end{eqnarray}
for all causally separable process matrices $W^\text{sep}$.
Since the set of causally separable process matrices is convex, then according to the separating hyperplane theorem~\cite{rockafellar70}, for any causally nonseparable $W^{\text{ns}}$ there must always exist a witness such that $\tr [S \cdot W^{\text{ns}} ] < 0$, which can thus be used to certify the causal nonseparability of $W^{\text{ns}}$; see Figure~\ref{fig:witness}. Note that the measurement of a witness is a \emph{device-dependent} test of causal nonseparability, as the physical operations of the parties must faithfully realise $S$ to be able to test Eq.~\eqref{eq:trSWgeq0}.

According to the above definition, and considering the trace as the Hilbert–Schmidt inner product, the set $\mathcal S$ of witnesses of causal nonseparability is simply the \emph{dual cone} (which we denote using an asterisk) of the cone of nonnormalised causally separable process matrices:
\begin{eqnarray}
\mathcal S &=& \big\{ \, S \, \big| \, \tr[S \cdot W^\text{sep}] \ge 0 \quad \forall \, W^\text{sep} \in \mathcal W^\text{sep} \big\} = (\mathcal W^\text{sep})^* \,. \nonumber \\
\end{eqnarray}
In the bipartite and particular tripartite cases considered here, this observation allows us to easily characterise the sets of witnesses $\mathcal S$, from the previous definitions of the corresponding cones $\wsepcone$; see Appendix~\ref{app:charact_cones}.

Note that for any $S^\perp$ in the orthogonal complement $\mathcal L_V^\perp$ of the linear subspace $\mathcal L_V$, and for any valid process matrix $W$ in $\mathcal W \subset \mathcal L_V$, one has $\tr[S^\perp \cdot W] = 0$. Hence, adding any term $S^\perp \in \mathcal L_V^\perp$ to a witness $S$ simply gives another witness, giving the same value of $\tr[S \cdot W]$ for any valid $W$. By choosing for instance $S^\perp = L_V(S) - S$, where $L_V$ is the projector onto the linear subspace $\mathcal L_V$, one thus obtains a witness in $\mathcal L_V$. For practical reasons, we will often be led to restrict the search of witnesses within the subspace $\mathcal L_V$; for that purpose we also define the (closed convex) cone of witnesses in $\mathcal L_V$ as $\mathcal S_V = \mathcal S \cap \mathcal L_V$.

\subsection{Determining causal (non)separability \newline through semidefinite programming}
\label{subsec:SDPs}

To determine whether a given process is causally separable or not, one possible approach is to rephrase the question as an optimisation problem, and ask how much noise can be added before it becomes causally separable.

Let us consider for now the case of `white noise', represented by the process matrix
\begin{equation}\label{white}
 \id^\circ = \frac{\id}{d_{I}} 
\end{equation}
with $d_I = d_{A_I}d_{B_I}$ or $d_I = d_{A_I}d_{B_I}d_{C_I}$ in the bipartite and tripartite cases,
and which prepares the incoming systems of all parties in a maximally mixed state.
For a given process matrix $W$ under consideration, we shall consider the noisy process
\begin{equation}
 W(r) = \frac{1}{1+r} \big( W + r \, \id^\circ \big)\,,
\end{equation}
and investigate its causal nonseparability.
Remembering that the normalisation of $ W(r)$ is irrelevant to check whether it is in the convex cone $\wsepcone$ of causally separable processes, this leads us to define the following optimisation problem:
\begin{equation}\label{eq:sdp_primal}
\begin{gathered}
 \min \ r \\
 \text{s.t.} \quad W + r \, \id^\circ \in \wsepcone \, .
\end{gathered}
\end{equation}

From the previous characterisation of the convex cone $\wsepcone$, one can see that this defines a semidefinite programming (SDP) problem~\cite{nesterov87}, which can be solved efficiently.
For ease of reference, we provide in Appendix~\ref{app:explicit_sdp} a more explicit description of this problem in terms of positive semidefinite constraints; see Eqs.~\eqref{eq:sdp_primal_explicit_bi} and~\eqref{eq:sdp_primal_explicit_tri} for the bipartite and tripartite cases, respectively. As can be seen, solving this problem provides an explicit decomposition of $W(r^*)$, where $r^*$ is the optimal solution of~\eqref{eq:sdp_primal}, as a convex combination of processes $W^{A \prec B}$ and $W^{B \prec A}$.
In analogy with the robustness of entanglement~\cite{vidal99}, the quantity $\max[r^*, 0]$ quantifies the robustness of the process $W$ with respect to white noise---or \emph{random robustness}~\cite{araujo15}.
In particular, a value $r^* > 0$ implies that $W$ is causally nonseparable.

The `primal' SDP problem~\eqref{eq:sdp_primal} is intimately linked to its `dual' problem, which is here~\cite{araujo15}
\begin{equation}\label{eq:sdp_dual}
\begin{gathered}
  \min \ \tr[S \cdot W] \\
 \text{s.t.} \quad S \in \mathcal S_V \mathand \tr[S \cdot \id^\circ] = 1 \, ,
\end{gathered}
\end{equation}
and whose optimal solution $S^*$ provides precisely, in the case where $\tr[S^* \cdot W] < 0$, a witness of the causal nonseparability of $W$.
Furthermore, the Duality Theorem for SDP problems~\cite{nesterov87} implies that the solutions of the primal and dual problems satisfy
\begin{equation} \label{eq:duality}
 r^* = -\tr[S^* \cdot W] \, .
\end{equation}
It follows in particular that $\tr[S^* \cdot W(r)] < 0$ for all $r < r^*$, i.e. for all $r$ such that $W(r)$ is causally nonseparable: this makes the witness $S^*$ optimal to detect the causal nonseparability of $W$ when subjected to white noise, see Figure~\ref{fig:witness}.

As for the primal problem, we provide in Appendix~\ref{app:explicit_sdp} a more explicit description of the dual problem~\eqref{eq:sdp_dual} that is better suited for practical use; see Eqs.~\eqref{eq:sdp_dual_explicit_bi} and~\eqref{eq:sdp_dual_explicit_tri}.
It is worth noting that, as discussed previously, adding any term $S^\perp \in \mathcal L_V^\perp$ to $S$ will not change the value of $\tr[S \cdot W]$, nor of $\tr[S \cdot \id^\circ]$. Hence, the problem~\eqref{eq:sdp_dual} is formally equivalent to one, where the constraint $S \in \mathcal S_V$ would be replaced by $S \in \mathcal S$; nevertheless, in practice, optimising over the whole (non-pointed) cone $\mathcal S$ may make the numerical solvers unstable~\cite{araujo15}.

\medskip

Note that depending on the practical physical implementation of a process $W$, different noise models may also be relevant. One could consider for instance a mixture with another fixed process $W^\circ$, and thus replace $\id^\circ$ in the primal SDP problem~\eqref{eq:sdp_primal} by $W^\circ$. The normalisation constraint in the dual problem~\eqref{eq:sdp_dual} would then be replaced by $\tr[S \cdot W^\circ] = 1$ and one can show, following similar proofs to those of Ref.~\cite{araujo15}, that as long as $W^\circ$ is in the relative interior of $\wsepcone$ (i.e., the interior of $\wsepcone$ within $\mathcal L_V$), the SDP problems would still be solved efficiently, with their optimal solutions still satisfying~\eqref{eq:duality}.

Another case of interest is that of robustness to \emph{worst case noise}, as also considered in Ref.~\cite{araujo15}.
One can define in this case the notion of \emph{generalised robustness} (again in analogy with entanglement~\cite{steiner03}), which can also be obtained through SDP. Interestingly, the generalised robustness can be used to define a proper \emph{measure} of causal nonseparability as it is (contrary to the random robustness) monotonous under local operations~\cite{araujo15}.

\subsection{Imposing further constraints on the witnesses}

In order to `measure' a witness $S$---i.e., to estimate the value $\tr[S \cdot W]$ (and check its sign)---one can in principle simply decompose it as a linear combination of products of CP (trace non-increasing) maps, implement these maps (provided this can be done even if the causal order between the parties is not well-defined), estimate their probabilities, and combine the statistics in an appropriate way (as illustrated for instance in the next section)~\cite{araujo15}.

In some cases, one may however not be able to implement all required CP maps, but may be restricted to CP maps from a certain class only---e.g., one may only be able to realise unitary operations.
In that case, not all witnesses can be measured, and it then makes sense to restrict the search of witnesses to those that are implementable in practice.
To do this, one can directly modify the dual problem~\eqref{eq:sdp_dual} and replace the search space $\mathcal S_V$ by the set $\tilde{\mathcal S} \subset \mathcal S$ of allowed witnesses (while no longer necessarily restricting the search to witnesses within $\mathcal L_V$).

Of course, with such an additional restriction the witnesses we shall obtain may not be optimal, and we will in general not be able to witness all causally nonseparable processes. Nevertheless, this possibility to add some constraints on the possible witnesses may be useful in practice, as we will illustrate below with the quantum switch.

\section{Case studies}

Let us now consider a few concrete examples to illustrate how one can construct witnesses and characterise causal nonseparability in practice. We start with a family of bipartite processes investigated already in Ref.~\cite{brukner14}, and then move on to the example of the quantum switch, for which we will consider different noise models and show how to add specific constraints on the witnesses we shall construct.

\subsection{A family of bipartite process matrices}
\label{subsec:W_etas}

In Ref.~\cite{brukner14}, the following family of process matrices was introduced:
\begin{eqnarray}
  W_{\eta_1, \eta_2} = \frac{1}{4} \Big[ \id &+& \eta_1 \, \id^{A_I} Z^{A_O} Z^{B_I} \id^{B_O} \nonumber \\[-2mm]
  &+& \eta_2 \, Z^{A_I} \id^{A_O} X^{B_I} Z^{B_O} \Big] \, , \label{eq:W_etas}
\end{eqnarray}
where $Z$ and $X$ are the Pauli matrices, the superscripts indicate to which system each operator is applied, and tensor products are implicit.
$W_{\eta_1, \eta_2}$ generalises in particular the process matrix originally considered in Ref.~\cite{oreshkov12}, obtained for $\eta_1 = \eta_2 = \frac{1}{\sqrt{2}}$.
One can easily check that $W_{\eta_1, \eta_2}$ satisfies Eqs.~\eqref{eq:valid_cond_bi_1BAA}--\eqref{eq:valid_cond_bi_1A1B} and~\eqref{eq:valid_cond_bi_norm}, and that it is positive semidefinite---hence, it is a valid process matrix---if and only if $\eta_1^2 + \eta_2^2 \leq 1$.

We solved, for different values of $\eta_1, \eta_2$, the dual SDP problem~\eqref{eq:sdp_dual}---or rather, its more explicit formulation given in~\eqref{eq:sdp_dual_explicit_bi}---using the Matlab software CVX~\cite{cvx}, and obtained (up to numerical precision) the witnesses
\begin{eqnarray}
  S_{\eta_1, \eta_2} = \frac{1}{4} \Big[ \id &-& \sgn(\eta_1) \, \id^{A_I} Z^{A_O} Z^{B_I} \id^{B_O} \nonumber \\[-2mm]
  &-& \sgn(\eta_2) \, Z^{A_I} \id^{A_O} X^{B_I} Z^{B_O} \Big] \, , \label{eq:S_etas}
\end{eqnarray}
where $\sgn$ is the sign function (for $\eta_1 = \eta_2 = \frac{1}{\sqrt{2}}$ we recover the witness obtained in Ref.~\cite{araujo15}).

To verify that $S_{\eta_1, \eta_2}$ is indeed a valid witness, one can check that ${}_{B_O}S_{\eta_1, \eta_2} \ge 0$ and ${}_{A_O}S_{\eta_1, \eta_2} \ge 0$: see the characterisation of witnesses in the bipartite case given in Appendix~\ref{app:charact_cones_S2}.
Applying $S_{\eta_1, \eta_2}$ to $W_{\eta_1, \eta_2}$, one gets
\begin{equation}
\tr[S_{\eta_1, \eta_2} \cdot W_{\eta_1, \eta_2}] = 1 - |\eta_1| - |\eta_2| \,,
\end{equation}
which shows that $W_{\eta_1, \eta_2}$ is causally nonseparable (the trace above is negative) for $|\eta_1| + |\eta_2| > 1$, and its random robustness in that case is $r^*_{\eta_1, \eta_2} = -\tr[S_{\eta_1, \eta_2} \cdot W_{\eta_1, \eta_2}] = |\eta_1| + |\eta_2| - 1$.

For $|\eta_1| + |\eta_2| \leq 1$ on the other hand, we find that $W_{\eta_1, \eta_2}$ is causally separable. Solving the primal SDP problem~\eqref{eq:sdp_primal}---or rather, its more explicit formulation~\eqref{eq:sdp_primal_explicit_bi}---provides an explicit decomposition as a convex sum of processes compatible with a definite causal order, in the form
\begin{eqnarray}
  W_{\eta_1, \eta_2} = \frac{|\eta_1|}{|\eta_1| {+} |\eta_2|} W_{\eta_1, \eta_2}^{A \prec B} + \frac{|\eta_2|}{|\eta_1| {+} |\eta_2|}  W_{\eta_1, \eta_2}^{B \prec A} \quad
\end{eqnarray}
with
\begin{eqnarray}
  W_{\eta_1, \eta_2}^{A \prec B} &=& \frac{1}{4} \Big[ \id + \sgn(\eta_1) \, \big(|\eta_1| {+} |\eta_2| \big) \, \id^{A_I} Z^{A_O} Z^{B_I} \id^{B_O} \Big] \, , \nonumber \\
  W_{\eta_1, \eta_2}^{B \prec A} &=& \frac{1}{4} \Big[ \id + \sgn(\eta_2) \, \big(|\eta_1| {+} |\eta_2| \big) \, Z^{A_I} \id^{A_O} X^{B_I} Z^{B_O} \Big] \, , \nonumber \\
\end{eqnarray}
where one can indeed check that $W_{\eta_1, \eta_2}^{A \prec B}$ and $W_{\eta_1, \eta_2}^{B \prec A}$ satisfy Eqs.~\eqref{eq:valid_cond_biAB_1B}--\eqref{eq:valid_cond_biAB_norm} and~\eqref{eq:valid_cond_biBA_1A}--\eqref{eq:valid_cond_biBA_norm}, as required (they are positive semidefinite precisely for $|\eta_1| + |\eta_2| \leq 1$).

\begin{figure}
 \includegraphics[width=\columnwidth]{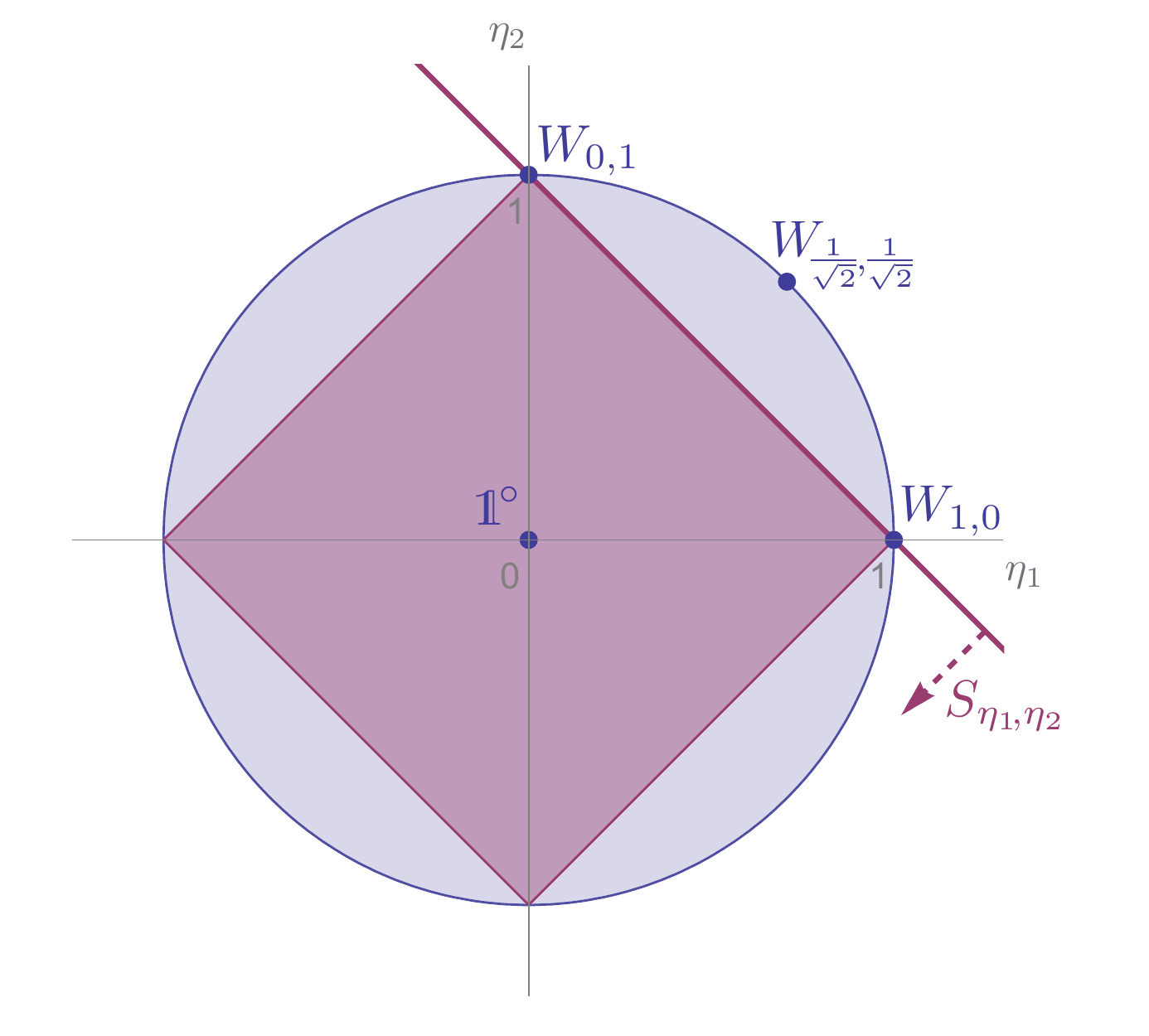}
\caption{Representation of the set of process matrices $W_{\eta_1, \eta_2}$ defined in Eq.~\eqref{eq:W_etas}. 
The shaded circle (characterised by $\eta_1^2 + \eta_2^2 \leq 1$) delimits the valid process matrices $W_{\eta_1, \eta_2} \ge 0$. Causally separable processes $W_{\eta_1, \eta_2}$ are restricted to the inner square ($|\eta_1| + |\eta_2| \leq 1$). Causally nonseparable processes (such that $|\eta_1| + |\eta_2| > 1$) can be witnessed by $S_{\eta_1, \eta_2}$~\eqref{eq:S_etas}, represented (for the case $\eta_1, \eta_2 \ge 0$) by the solid line. The figure here is similar to Figure~2 of Ref.~\cite{brukner14}.}
\label{fig:W_etas}
\end{figure}

Figure~\ref{fig:W_etas} represents the set of process matrices $W_{\eta_1, \eta_2}$. We recover here the results found in Ref.~\cite{brukner14}; however, the use of witnesses allows us to give a much more direct proof of causally (non)separability for the $W_{\eta_1, \eta_2}$ matrices.

\medskip

In order to measure the witness $S_{\eta_1, \eta_2}$ in practice, one can for instance decompose its two nontrivial components in terms of CP (trace non-increasing) maps as follows:
\begin{subequations}
\begin{eqnarray}
  \id^{A_I} Z^{A_O} Z^{B_I} \id^{B_O} & = & 4 \! \sum_{a,b = \pm 1} a \, b \ M_{a|\id Z}^{A_IA_O} \!\otimes\! M_{b|Z \id}^{B_IB_O} , \\
  Z^{A_I} \id^{A_O} X^{B_I} Z^{B_O} & = & 2 \!\! \sum_{\tau,a,b = \pm 1} \!\! \tau \, a \, b \ M_{a|Z \id}^{A_IA_O} \!\otimes\! M_{b|X, \tau Z}^{B_IB_O} \, , \quad \nonumber \\[-4mm]
\end{eqnarray}
\end{subequations}
with
\begin{eqnarray}
  & M_{\pm|\id Z} = \frac{\id}{2} \otimes \frac{\id \pm Z}{2} \, , \quad
  M_{\pm|Z \id} = \frac{\id \pm Z}{2} \otimes \frac{\id}{2} \, , \nonumber \\
  & M_{\pm|X,\tau Z} = \frac{\id \pm X}{2} \otimes \frac{\id + \tau Z}{2}
\end{eqnarray}
(where the second part of the subscripts denote a particular choice of instrument: a choice of `setting'),
and then calculate, using the generalised Born rule~\eqref{born},
\begin{eqnarray}
  \tr[S_{\eta_1, \eta_2} \! \cdot \! W] &=& 1 - \sgn(\eta_1) \, \sum_{a,b} a \, b \ P( M_{a|\id Z}^{A_IA_O}, M_{b|Z \id}^{B_IB_O}) \nonumber \\[-1mm]
  &&\, - \sgn(\eta_2) \frac{1}{2} \!\sum_{\tau,a,b} \tau \, a \, b \ P( M_{a|Z \id}^{A_IA_O} \! , M_{b|X, \tau Z}^{B_IB_O} ) \, . \nonumber \\[-3mm]
\end{eqnarray}
(Note that the decomposition of a witness in terms of CP maps is not unique; another possible decomposition of $S_{\eta_1, \eta_2}$, for the case $\eta_1, \eta_2 >0$, was given in Ref.~\cite{araujo15}.)

\subsection{The quantum switch}
\label{subsec_Wswitch}

The quantum switch is a circuit, which was proposed to extend the framework of causally ordered quantum circuits and allow the order in which gates are performed to be coherently controlled by a quantum system~\cite{chiribella09}. As proven recently~\cite{araujo15,oreshkov15}, when analysed in the framework of process matrices, the quantum switch provides precisely an example of a (tripartite) causally nonseparable process. It is in fact the first practical example that we know how to realise physically (and which has been demonstrated experimentally~\cite{procopio_experimental_2014}), as, to the best of our knowledge, no practical realisation is known so far for any of the causally nonseparable process matrices exhibited, e.g., in Refs.~\cite{oreshkov12,baumeler13, baumeler14,brukner14,branciard16}.

In its simplest version, the quantum switch involves two qubits---a control qubit and a target qubit. The target qubit, initially prepared in some state $\ket{\psi}$, is sent to two parties, Alice and Bob, who act on it in an order that is determined by the state of the control qubit: if the control qubit is in the state $\ket{0}$, then Alice acts first and Bob acts second, while if it is in the state $\ket{1}$, then Bob acts first and Alice second. The interesting situation is when the control qubit is in a superposition $\frac{1}{\sqrt{2}}(\ket{0}+\ket{1})$, in which case Alice and Bob can be said to act `in a superposition of orders'. After Alice and Bob's operations, the control qubit is sent to a third party, Charlie, who can measure it.

As shown in Ref.~\cite{araujo15} (see also ~\cite{oreshkov15}), the quantum switch can be represented in terms of the `pure process'
\begin{eqnarray}
\label{eq:qswitch_ketw}
\ket{w} &=&  \frac{1}{\sqrt{2}}\left(\ket{\psi}^{A_I}\Ket{\id}^{A_OB_I}\Ket{\id}^{B_OT_I}\ket{0}^{C_I} \right. \nonumber \\[-2mm]
&& \qquad \ + \left. \ket{\psi}^{B_I}\Ket{\id}^{B_OA_I}\Ket{\id}^{A_OT_I}\ket{1}^{C_I} \right),
\end{eqnarray}
where $\Ket{\id} = \ket{00}+\ket{11}$ is the CJ representation of an identity qubit channel. After tracing out the target qubit in the system $T_I$, we obtain the process matrix representing the quantum switch as\footnote{Alternatively, the target qubit could also be sent for instance to Charlie, would could measure it together with the control qubit; for simplicity we do not consider this possibility here.}
  \begin{equation}\label{eq:qswitch_W}
   W_\text{switch} = \tr_{T_I} \proj{w} \, .
  \end{equation}
Note that $W_\text{switch} \in A_I\otimes A_O \otimes B_I\otimes B_O \otimes C_I$ (with $d_{A_I} = d_{A_O} = d_{B_I} = d_{B_O} = d_{C_I} = 2$) and that Charlie has no output system, so that we are indeed in the particular tripartite case considered previously.

\subsubsection{Robustness to white noise}

To investigate the causal nonseparability of the quantum switch and construct a witness, one can follow the approach described in Subsection~\ref{subsec:SDPs}. We solved the SDP problems~\eqref{eq:sdp_primal}--\eqref{eq:sdp_dual}---or rather, their more explicit formulation~\eqref{eq:sdp_primal_explicit_tri}--\eqref{eq:sdp_dual_explicit_tri}---numerically with CVX~\cite{cvx}, and found that the random robustness of the quantum switch is
\begin{equation}
r^*_\text{switch} \simeq 1.576 \,.
\end{equation}
Alternatively, in terms of the `visibility' $v$, this means that the noisy quantum switch
\begin{equation}\label{eq:qswitch_W_white}
   W_\text{switch}^{\id^{\!\circ}}(v) = v \, W_\text{switch} + (1{-}v) \, \id^\circ
\end{equation}
is causally nonseparable for all $v \ge v^*_\text{switch} = \frac{1}{1+r^*_\text{switch}} \simeq 0.3882$.
The explicit witness $S_\text{switch}$ obtained numerically from the dual SDP problem~\eqref{eq:sdp_dual} is given in Appendix~\ref{app:S_switch}.

\subsubsection{Depolarising the control qubit}

In a practical implementation of the quantum switch, other noise models than fully white noise can also be relevant.

Consider for instance a situation where, for practical reasons, the target qubit is well preserved throughout the setup, but the control qubit is affected by white noise: with some probability $v$ (which can be understood as a `visibility'), the state of the control qubit is untouched, and with some probability $1-v$ it is depolarised to the fully fixed state $\id^{C_I}/2$. The resulting noisy process then writes
\begin{eqnarray}
   W_\text{switch}^\text{depol}(v) &=& v \, W_\text{switch} + (1{-}v) \, W_\text{depol}
\label{eq:qswitch_W_depol}
\end{eqnarray}
with
\begin{eqnarray}
   W_\text{depol} &=& {}_{C_I}W_\text{switch} \nonumber \\
   &=& \frac{1}{2}\left( \ketbra{\psi}{\psi}^{A_I} \!\otimes \KetBra{\id}^{A_OB_I} \!\otimes \id^{B_O} \right. \nonumber \\[-2mm]
&& \ + \left. \ketbra{\psi}{\psi}^{B_I} \!\otimes \KetBra{\id}^{B_OA_I} \!\otimes \id^{A_O} \right) \otimes \frac{\id^{C_I}}{2} \,, \qquad
\end{eqnarray}
which corresponds to a random mixture of a process where the target qubit goes first to Alice then to Bob, and a process where it goes first to Bob and then to Alice.

One clearly sees that $W_\text{depol}$ is causally separable.
As it turns out, it lies precisely on the boundary of the set of causally separable processes; hence, some care needs to be taken if one wants to investigate the causal (non)separability of $W_\text{switch}^\text{depol}(v)$ as discussed at the end of Subsection~\ref{subsec:SDPs}. A possible approach is for instance to mix the quantum switch with a process that is $\epsilon$-close to $W_\text{depol}$ (and let $\epsilon \to 0$), inside the relative interior of $\wsepcone$; or to directly calculate the random robustness of $W_\text{switch}^\text{depol}(v)$ for various fixed values of $v$.

By doing so, we found numerically a positive random robustness for all chosen values $v > 0$. In fact, one can prove analytically that $W_\text{switch}^\text{depol}(v)$ is causally nonseparable whenever $v>0$, by constructing a family of witnesses $S(v)$ such that $\tr[S(v) \cdot W_\text{switch}^\text{depol}(v)] < 0$ for all $v > 0$; see Appendix~\ref{app:S_v}. That is, the causal nonseparability of the quantum switch is infinitely robust to white noise affecting the control qubit only.

\begin{figure}
 \includegraphics[width=\columnwidth]{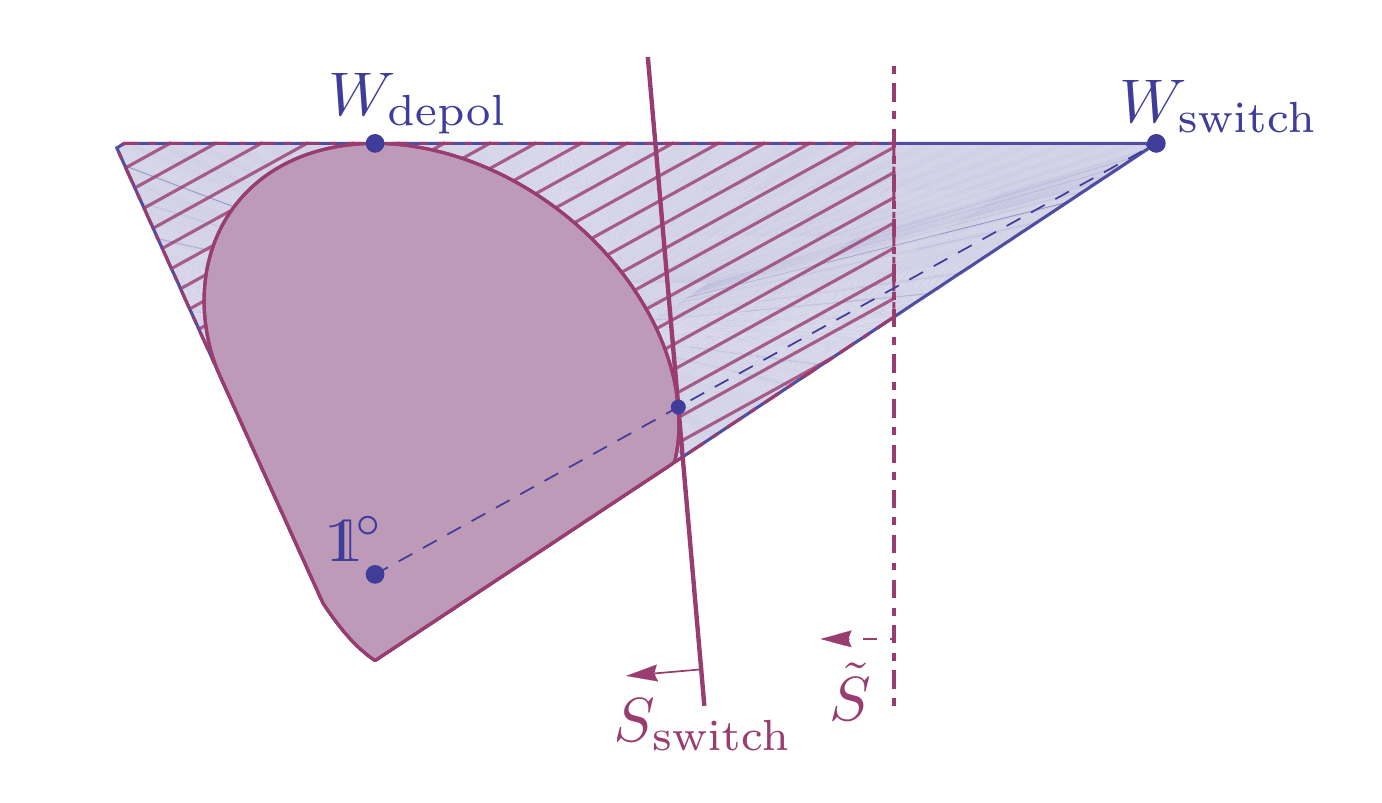}
\caption{Two-dimensional slice of the space of process matrices containing $W_\text{switch}, W_\text{depol}$ and $\id^\circ$. The shaded region contains all valid (positive semidefinite) process matrices, with the inner darker region containing the causally separable processes.
The causal nonseparability of $W_\text{switch}$ can be witnessed using $S_\text{switch}$, given explicitly in Appendix~\ref{app:S_switch}, which is optimal to test its robustness to white noise. All processes $W_\text{switch}^\text{depol}(v) = v \, W_\text{switch} + (1{-}v) \, W_\text{depol}$ with $0 < v \leq 1$ are causally nonseparable, as can be shown using a family of witnesses given in Appendix~\ref{app:S_v}. The witness $\tilde S$ can be measured with Alice and Bob restricting their operations to unitaries; only the causally nonseparable processes outside of the hatched region can be witnessed with this restriction.}
\label{fig:W_depol}
\end{figure}

Figure~\ref{fig:W_depol} shows, for illustration, the two-dimensional slice of the space of process matrices that contains $W_\text{switch}, W_\text{depol}$ and $\id^\circ$. By scanning this whole slice, one can characterise using our SDP technique the limits of the set of causally separable processes. One can clearly see for instance that the whole line segment containing the processes $W_\text{switch}^\text{depol}(v)$ with $v>0$ is outside of it, and approaches it tangentially.

\subsubsection{Dephasing the control qubit}

Rather than fully depolarising the control qubit, it may be relevant to investigate the case where it is only dephased, i.e. it undergoes (with some probability $1-v$, as before) the map
\begin{equation}
\rho \to \bra{0}\rho\ket{0} \, \ketbra{0}{0} + \bra{1}\rho\ket{1} \, \ketbra{1}{1} \,,
\end{equation}
so that its coherence is lost.

We are thus led to consider here the noisy process
\begin{eqnarray}
   W_\text{switch}^\text{deph}(v) &=& v \, W_\text{switch} + (1{-}v) \, W_\text{deph}
\label{eq:qswitch_W_deph}
\end{eqnarray}
with
\begin{eqnarray}
   W_\text{deph} &=& \frac{1}{2}\left( \ketbra{\psi}{\psi}^{A_I} \!\otimes \KetBra{\id}^{A_OB_I} \!\otimes \id^{B_O} \!\otimes \ketbra{0}{0}^{C_I} \right. \nonumber \\[-2mm]
&& \ + \left. \ketbra{\psi}{\psi}^{B_I} \!\otimes \KetBra{\id}^{B_OA_I} \!\otimes \id^{A_O} \!\otimes \ketbra{1}{1}^{C_I} \right) , \qquad
\end{eqnarray}
which corresponds now to a situation where a classical control bit, in the state $\ketbra{0}{0}^{C_I}$ or $\ketbra{1}{1}^{C_I}$ with equal probability, determines the order between Alice and Bob---a process that we could call a \emph{classical switch}.

\begin{figure}
 \includegraphics[width=\columnwidth]{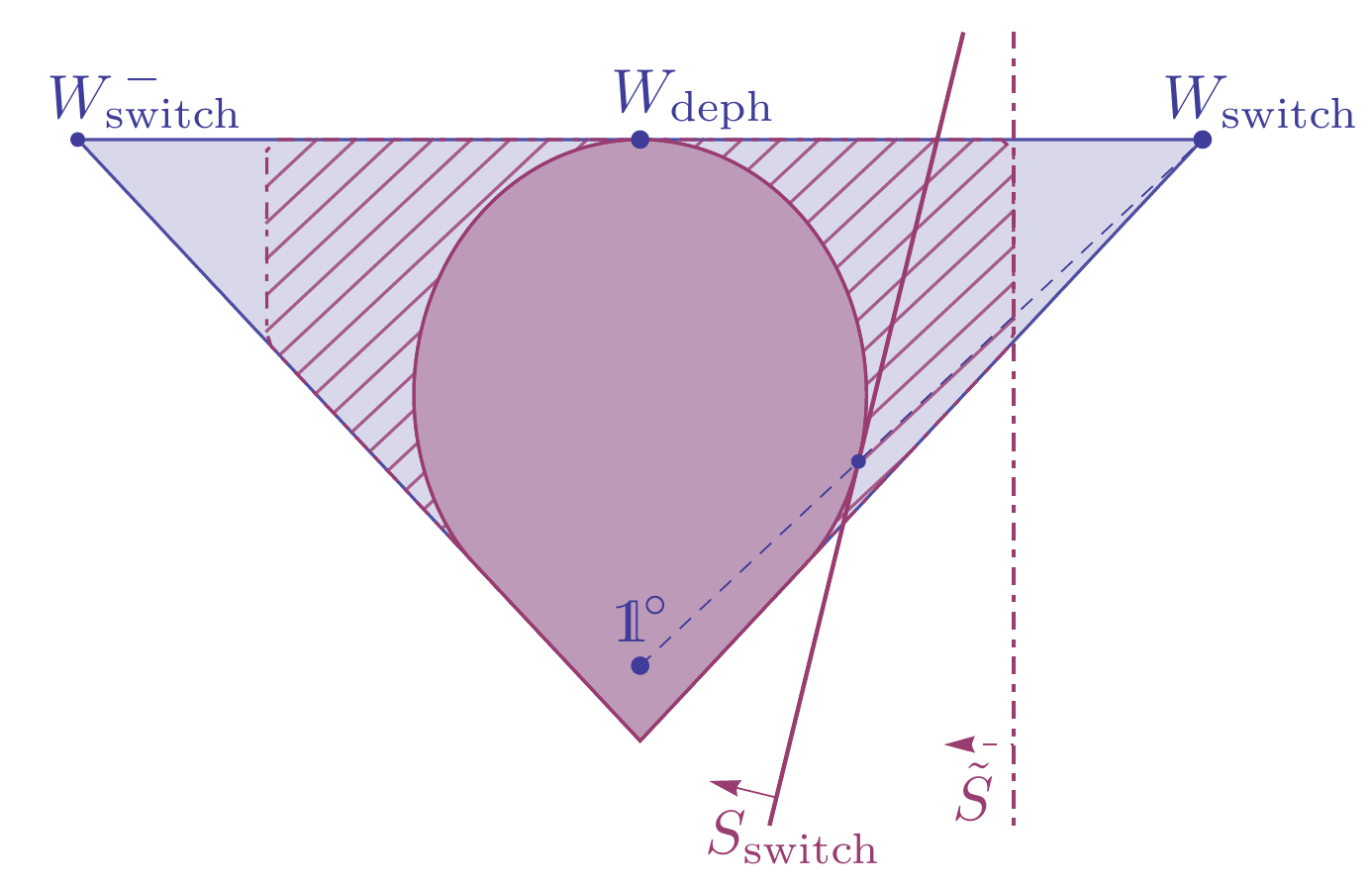}
\caption{Analogous figure to Fig.~\ref{fig:W_depol}, for the two-dimensional slice of the space of process matrices containing now $W_\text{switch}, W_\text{deph}$ and $\id^\circ$. The process $W_\text{switch}^-$, symmetric to $W_\text{switch}$, is the process obtained when implementing the quantum switch with a control qubit initially in the state $\frac{1}{\sqrt{2}}(\ket{0}-\ket{1})$ rather than $\frac{1}{\sqrt{2}}(\ket{0}+\ket{1})$ (whose description as a process matrix is then obtained by replacing the `$+$' sign by a `$-$' sign in Eq.~\eqref{eq:qswitch_ketw}).}
\label{fig:W_deph}
\end{figure}

Clearly, $W_\text{deph}$ is causally separable. Like $W_\text{depol}$, it also lies on the boundary of the set of causally separable processes. One can again check numerically and prove analytically (see Appendix~\ref{app:S_v}) that $W_\text{switch}^\text{deph}(v)$ is causally nonseparable for all $v>0$: that is, the quantum switch is also infinitely robust to dephasing noise affecting the control qubit only.
As with Figure~\ref{fig:W_depol}, Figure~\ref{fig:W_deph} now shows, for illustration, the two-dimensional slice of the space of process matrices that contains $W_\text{switch}, W_\text{deph}$ and $\id^\circ$.

\subsubsection{Restricting Alice and Bob's operations to unitaries}

To finish with, let us consider an implementation of the quantum switch where Alice and Bob are restricted to perform unitary operations. This restriction is motivated by practical reasons: in the recent photonic implementation of the quantum switch reported in Ref.~\cite{procopio_experimental_2014} for example, Alice and Bob only used passive optical elements, namely half  and quarter wave plates, realising (up to experimental imperfections) unitaries on the target qubit, encoded in the photon polarisation. In particular, Alice and Bob do not perform any actual measurement, and do not need to record measurement outcomes (only Charlie makes a measurement with different possible outcomes).

As we show in Appendix~\ref{app:S_U}, the CJ representation $M_U^{X_IX_O}$ of a unitary operation $U : \mathcal H^{X_I} \to \mathcal H^{X_O}$ satisfies
\begin{eqnarray}
{}_{X_I}M_U^{X_IX_O} = {}_{X_O}M_U^{X_IX_O} = {}_{X_IX_O}M_U^{X_IX_O} \,. \label{eq:constr_U}
\end{eqnarray}
Now, if Alice and Bob are restricted to perform unitary operations, the witnesses that can be measured must be of the form 
\begin{equation}\label{eq:Witn3_Decomp_unitaries}
S = \sum_{x,y,z,c} \gamma_{x,y,z,c} \ M_{U_x}^{A_IA_O} \otimes M_{U_y}^{B_IB_O} \otimes M_{c|z}^{C_I} \, ,
\end{equation}
for some unitaries $U_x, U_y$, for some CP maps (or simply: POVM elements) $M_{c|z}^{C_I}$, and some real coefficients $\gamma_{x,y,z,c}$. Because of~\eqref{eq:constr_U}, $S$ will then necessarily satisfy
\begin{eqnarray}
{}_{A_I}S = {}_{A_O}S = {}_{A_IA_O}S \ \ \text{and} \ \ {}_{B_I}S = {}_{B_O}S = {}_{B_IB_O}S \,. \label{eq:constr_U_S} \qquad
\end{eqnarray}

Hence, to construct such a witness, one can simply solve the dual SDP problem~\eqref{eq:sdp_dual}, replacing the constraint $S \in \mathcal S_V$ by $S \in \tilde{\mathcal S}$, with
\begin{eqnarray}
\tilde{\mathcal S} = \big\{ \, S \in \mathcal S \, \big| \, {}_{A_I}S = {}_{A_O}S = {}_{A_IA_O}S \ \qquad \nonumber \\[-1mm]
\text{and} \ {}_{B_I}S = {}_{B_O}S = {}_{B_IB_O}S \big\} \,.
\end{eqnarray}
The resulting optimisation problem remains a SDP problem. Solving it with CVX, we obtained numerically an explicit witness $\tilde S$, given in Appendix~\ref{app:S_U} and shown on Figures~\ref{fig:W_depol} and~\ref{fig:W_deph}, that detects the causal nonseparability of the processes $W_\text{switch}^{\id^{\!\circ}}(v)$~\eqref{eq:qswitch_W_white}, $W_\text{switch}^\text{depol}(v)$~\eqref{eq:qswitch_W_depol} and $W_\text{switch}^\text{deph}(v)$~\eqref{eq:qswitch_W_deph} down to $v \simeq 0.6641$ (the same value for all three).

Clearly, the price to pay by restricting Alice and Bob to unitaries only is that not all causally nonseparable processes can be witnessed; see the hatched regions in Figures~\ref{fig:W_depol} and~\ref{fig:W_deph}. Nevertheless, the amount of noise tolerated by $\tilde S$ is already good enough to measure it and demonstrate causal nonseparability experimentally with current technologies, e.g. in a setup similar to that of Ref.~\cite{procopio_experimental_2014}.

\section{Discussion}

In this paper we have given an introduction to witnesses of causal nonseparability~\cite{araujo15}, and illustrated this concept on a few explicit examples.
Witnesses of causal nonseparability are somewhat analogous to entanglement witnesses; however, a remarkable difference is that contrary to the latter, the former can be constructed efficiently, for any causally nonseparable processes (in the bipartite and particular tripartite cases considered here), using semidefinite programming.

Among the explicit examples given above, of particular interest is the quantum switch. This is indeed the first concrete example of a causally nonseparable process that we know how to realise in practice, and for which we know how to witness the causal nonseparability.
We constructed its optimal witness with respect to white noise, which detects its causal nonseparability down to a visibility of $v \simeq 0.3882$. We further constructed a witness that can be measured with Alice and Bob implementing unitaries only, and which is robust to visibilities down to $v \simeq 0.6441$---whether we consider white noise, or depolarising or dephasing noise that affects the control qubit only. This allows for a feasible experimental verification of the causal nonseparability of the quantum switch 
that would be more robust than with the witness previously proposed in~\cite{araujo15}, which allows only for visibilities down to $v \simeq 0.7381$ (corresponding to a success probability $p_\text{succ} = \frac{1+v}{2} \simeq 0.8690$ for Chiribella's task~\cite{chiribella12}, as reported in~\cite{araujo15}).
Note that in the latter, Charlie only performs measurements in the $X$ basis (while our witness also involves the $Y$ basis, see Appendix~\ref{app:S_U}); as it turns out, that witness was actually optimal under this restriction, as can be shown by further adding the corresponding constraint in the dual problem~\eqref{eq:sdp_dual}. 
Recall that the witness obtained in~\cite{araujo15} was constructed from Chiribella's task of distinguishing between a commuting and an anticommuting channel, where the quantum switch provides an advantage over any causally ordered circuit~\cite{chiribella12}.
We note indeed that the tool of witnesses of causal nonseparability and the techniques developed to construct them may also be useful to inspire and analyse possible applications of causally nonseparable processes~\cite{chiribella12,araujo14,feix15}, and to quantify their advantages over causally separable resources.

Let us finish by emphasising that in this paper, as in Ref.~\cite{araujo15}, we only considered the bipartite case and a particular tripartite case, where the third party has no (or a trivial) outgoing system. Characterising and constructing witnesses in the general case remains so far an open problem. Clearly, the sets of nonnormalised process matrices and of witnesses remain closed convex cones, and one can still write the optimisation problems~\eqref{eq:sdp_primal} and~\eqref{eq:sdp_dual} as conic problems. However, whether the characterisation of the cones $\wsepcone$ and $\mathcal S_{(V)}$ would allow us to write them as SDP problems that can be solved efficiently, and whether the duality relation~\eqref{eq:duality} would still hold, is left for future research.

\subsection*{Acknowledgements}

I acknowledge fruitful discussions with all my co-authors of Ref.~\cite{araujo15} and feedback on this manuscript from Alastair Abbott. This work was funded by the `Retour Post-Doctorants' program (ANR-13-PDOC-0026) of the French National Research Agency and by a Marie Curie International Incoming Fellowship (PIIF-GA-2013-623456) of the European Commission.

\appendix

\section{Characterisation of the cones \newline $\mathcal W^\text{sep}$, $\mathcal S$ and $\mathcal S_V$}
\label{app:charact_cones}

In this Appendix we show how to explicitly characterise the cones $\mathcal W^\text{sep}$ of (nonnormalised) causally separable process matrices, and the cones $\mathcal S = (\mathcal W^\text{sep})^*$ and $\mathcal S_V = \mathcal S \cap \mathcal L_V$ of witnesses of causal nonseparability, in the bipartite and particular tripartite cases considered in the main text.

The characterisations below were first obtained in Ref.~\cite{araujo15}. In what follows here, it is implicit that all matrices under consideration are either in $A_I\otimes A_O\otimes B_I\otimes B_O$ (in the bipartite case) or $A_I\otimes A_O\otimes B_I\otimes B_O \otimes C_I$ (in the tripartite case); in particular, they are all Hermitian.

\subsection{$\mathcal W^\text{sep}$: Causally separable process matrices}

\subsubsection{Bipartite case}

According to Eq.~\eqref{eq:Wsep_cone_bi}, bipartite causally separable process matrices can be written as
\begin{equation}
W = W^{A \prec B} + W^{B \prec A} \label{eq:Wsep_decomp_bi}
\end{equation}
with $W^{A \prec B}$ and $W^{B \prec A}$ two positive semidefinite matrices satisyfing~\eqref{eq:valid_cond_biAB_1B}--\eqref{eq:valid_cond_biAB_1ABB} and~\eqref{eq:valid_cond_biBA_1A}--\eqref{eq:valid_cond_biBA_1BAA}, respectively.

Note that if $W$ is already assumed to be a valid process matrix in $\mathcal L_V$ (hence, it satisfies in particular Eqs.~\eqref{eq:valid_cond_bi_1BAA} and~\eqref{eq:valid_cond_bi_1ABB}), then assuming that $W^{A \prec B}$ satisfies~\eqref{eq:valid_cond_biAB_1B} automatically implies that $W^{B \prec A} = W - W^{A \prec B}$ satisfies~\eqref{eq:valid_cond_biBA_1BAA}; similarly, assuming that $W^{B \prec A}$ satisfies~\eqref{eq:valid_cond_biBA_1A} automatically implies that $W^{A \prec B}$ satisfies~\eqref{eq:valid_cond_biAB_1ABB}. Hence, to determine whether $W \in \mathcal L_V$ is causally separable, it is enough to check whether it can be decomposed as in~\eqref{eq:Wsep_decomp_bi} with $W^{A \prec B} \ge 0$ and $W^{B \prec A} \ge 0$ satisfying~\eqref{eq:valid_cond_biAB_1B} and~\eqref{eq:valid_cond_biBA_1A}, resp. Defining the linear subspaces
\begin{eqnarray}
\mathcal L_{[1-B_O]} &=& \{ W \,|\, {}_{[1-B_O]}W = 0 \} \,, \\
\mathcal L_{[1-A_O]} &=& \{ W \,|\, {}_{[1-A_O]}W = 0 \} \,,
\end{eqnarray}
the cone of (nonnormalised) causally separable process matrices can then be characterised as~\cite{araujo15}
\begin{eqnarray}
\wsepcone &=& {\cal W}^{A \prec B} + {\cal W}^{B \prec A} \\
&=& \big[ (\mathcal P \cap \mathcal L_{[1-B_O]}) + (\mathcal P \cap \mathcal L_{[1-A_O]}) \big] \cap \mathcal L_V \,. \label{eq:Wsep_cone_bi_v2} \quad
\end{eqnarray}

\subsubsection{Tripartite case with $d_{C_O} = 1$}

With similar arguments as in the bipartite case above, we find that in the particular tripartite case where Charlie has a trivial outgoing system ($d_{C_O} = 1$), the cone of (nonnormalised) causally separable process matrices can be characterised as
\begin{eqnarray}
\wsepcone &=& {\cal W}^{A \prec B \prec C} + {\cal W}^{B \prec A \prec C} \label{eq:Wsep_cone_tri_v1} \\
&=& \big[ (\mathcal P \cap \mathcal L_{[1-B_O]C_I}) \nonumber \\[-1mm]
&& \hspace{3mm} + (\mathcal P \cap \mathcal L_{[1-A_O]C_I}) \big] \cap \mathcal L_V \quad \label{eq:Wsep_cone_tri_v2}
\end{eqnarray}
with
\begin{eqnarray}
\mathcal L_{[1-B_O]C_I} &=& \{ W \,|\, {}_{[1-B_O]C_I}W = 0 \} \,, \\
\mathcal L_{[1-A_O]C_I} &=& \{ W \,|\, {}_{[1-A_O]C_I}W = 0 \} \,.
\end{eqnarray}

\subsection{$\mathcal S$ and $\mathcal S_V$: Witnesses of causal nonseparability}

As explained in the main text, the set of witnesses of causal nonseparability is simply the dual cone of $\wsepcone$. It can be characterised by using the previous descriptions of $\wsepcone$, and making use of the following duality relations for two nonempty closed convex cones $\mathcal K_1, \mathcal K_2$~\cite{rockafellar70}:
\begin{equation}
(\mathcal K_1 + \mathcal K_2 )^* = \mathcal K_1^* \cap \mathcal K_2^* \,, \quad (\mathcal K_1 \cap \mathcal K_2 )^* = \mathcal K_1^* + \mathcal K_2^* \,. \ \label{eq:duality_relations}
\end{equation}

\subsubsection{Bipartite case}
\label{app:charact_cones_S2}

Using~\eqref{eq:Wsep_cone_bi_v2} and~\eqref{eq:duality_relations}, noting that the dual cone of a linear subspace $\mathcal L$ is its orthogonal complement  $\mathcal L^\perp$
and that the cone $\mathcal P$ of positive semidefinite matrices is self-dual, one can write, in the bipartite case,
\begin{eqnarray}
\mathcal S &=& (\wsepcone)^* \\
&=& \big[ (\mathcal P + \mathcal L_{[1-B_O]}^\perp) \cap (\mathcal P + \mathcal L_{[1-A_O]}^\perp) \big] + \mathcal L_V^\perp \,. \qquad \label{eq:S_cone_bi}
\end{eqnarray}

Noting now that $\mathcal L_{[1-B_O]}^\perp = \{ S \,|\, {}_{[1-B_O]}S = S \} = \{ S \,|\, {}_{B_O}S = 0 \}$ and that the map $S \to {}_{B_O}S$ is positive, one can easily show~\cite{araujo15} that $\mathcal P + \mathcal L_{[1-B_O]}^\perp = \{ S \,|\, {}_{B_O}S \ge 0 \}$, and similarly that $\mathcal P + \mathcal L_{[1-A_O]}^\perp = \{ S \,|\, {}_{A_O}S \ge 0 \}$.
Furthermore, one has $\mathcal L_V^\perp = \{S | L_V(S) = 0 \}$, where $L_V$ is the projector onto the linear subspace $\mathcal L_V (= \{S | L_V(S) = S \})$, which can be written as
\begin{eqnarray}
&& \hspace{-2mm} L_V(W) \nonumber \\
&& = {}_{\big[ 1 - [1-B_O]A_IA_O \big] \big[ 1 - [1-A_O]B_IB_O \big] \big[ 1 - [1-A_O][1-B_O] \big]}W \quad \nonumber \\
&& = {}_{\big[ 1 - [1-B_O]A_IA_O - [1-A_O]B_IB_O - [1-A_O][1-B_O] \big]}W . \nonumber \\[-5mm]
 \label{def:LV2}
\end{eqnarray}

Combining this with~\eqref{eq:S_cone_bi}, we find that
\begin{eqnarray}
{\cal S} \, = \, \Big\{ S = S^P + S^\perp \, \Big| \, {}_{B_O}S^P \geq 0 \, , \ {}_{A_O}S^P \geq 0 \, , \quad \ \nonumber \\[-2mm]
L_V(S^\perp) = 0 \, \Big\} . && \qquad \quad
\label{eq:S2_characterisation}
\end{eqnarray}

Furthermore, $S = S^P + S^\perp$ thus characterised is in $\mathcal L_V$ if and only if $S = L_V(S_P + S^\perp) = L_V(S_P)$. Hence, we also simply have
\begin{eqnarray}
{\cal S}_V \, = \, \Big\{ S = L_V(S^P) \, \Big| \, {}_{B_O}S^P \geq 0 \, , \ {}_{A_O}S^P \geq 0 \, \Big\} \, . \qquad \quad
\label{eq:SV2_characterisation}
\end{eqnarray}

\subsubsection{Tripartite case with $d_{C_O} = 1$}
\label{app:charact_cones_S3}

One could follow a similar reasoning as above to characterise $\mathcal S$ in the tripartite case with $d_{C_O} = 1$, starting from the characterisation of $\wsepcone$ given by Eq.~\eqref{eq:Wsep_cone_tri_v2}. However, because the map $S \to {}_{[1-(1-B_O)C_I]}S$ is not positive (contrary to $S \to {}_{[1-(1-B_O)]}S = {}_{B_O}S$), one cannot simplify the characterisation of $\mathcal P + \mathcal L_{[1-B_O]C_I}^\perp$---and ultimately of $\mathcal S$---as much as before.

It is thus somewhat simpler here to start directly from the characterisation of $\wsepcone$ given by Eq.~\eqref{eq:Wsep_cone_tri_v1}. With ${\cal W}^{A \prec B \prec C} = \mathcal P \cap \mathcal L_{A \prec B \prec C}$ and ${\cal W}^{B \prec A \prec C} = \mathcal P \cap \mathcal L_{B \prec A \prec C}$, we get, using again the relations~\eqref{eq:duality_relations},
\begin{eqnarray}
\mathcal S &=& (\wsepcone)^* \nonumber \\
&=& \big[ (\mathcal P \cap \mathcal L_{A \prec B \prec C}) + (\mathcal P \cap \mathcal L_{B \prec A \prec C}) \big]^* \nonumber \\
&=& (\mathcal P + \mathcal L_{A \prec B \prec C}^\perp) \cap (\mathcal P + \mathcal L_{B \prec A \prec C}^\perp) \nonumber \\[2mm]
&=& \Big\{ \, S = S_{ABC}^P + S_{ABC}^\perp = S_{BAC}^P + S_{BAC}^\perp \nonumber \\[-1mm]
&& \qquad \Big| \ S_{ABC}^P \geq 0 \, , \ L_{A \prec B \prec C}(S_{ABC}^\perp) = 0 \, , \nonumber \\[-2mm]
&& \qquad \ \ S_{BAC}^P \geq 0 \, , \ L_{B \prec A \prec C}(S_{ABC}^\perp) = 0 \ \Big\} \, , \qquad \label{eq:S_tri_characterisation}
\end{eqnarray}
where $L_{A \prec B \prec C}$ and $L_{B \prec A \prec C}$ are the projectors onto the linear subspaces $\mathcal L_{A \prec B \prec C}$ and $\mathcal L_{B \prec A \prec C}$, which are
\begin{eqnarray}
L_{A \prec B \prec C}(W) &=& {}_{\big[ 1 - [1-B_O]C_I - [1-A_O]B_IB_OC_I \big]}W \,, \qquad \label{def:LABC} \\
L_{B \prec A \prec C}(W) &=& {}_{\big[ 1 - [1-A_O]C_I - [1-B_O]A_IA_OC_I \big]}W \,. \label{def:LBAC}
\end{eqnarray}

Restricting the witnesses to the subspace $\mathcal L_V$, one can then write
\begin{equation}
{\cal S}_V = \{\, S \in {\cal S} \, | \, L_V(S) = S \,\}
\label{eq:SV3_characterisation}
\end{equation}
by referring to the previous characterisation~\eqref{eq:S_tri_characterisation} of $\mathcal S$, and with the projector $L_V$ onto $\mathcal L_V$ now given by
\begin{eqnarray}
L_V(W) = {}_{\big[ 1 - [1-B_O]A_IA_OC_I - [1-A_O]B_IB_OC_I} \qquad \qquad \quad \nonumber \\[-3mm]
 {}_{- [1-A_O][1-B_O]C_I \big]}W \,. \qquad \quad
 \label{def:LV3}
\end{eqnarray}

\section{Explicit formulation \newline of our SDP problems}
\label{app:explicit_sdp}

The previous characterisations of the cones $\wsepcone$ and $\mathcal S_V$ allow us to write (in our bipartite and tripartite cases) the primal and dual SDP problems~\eqref{eq:sdp_primal} and~\eqref{eq:sdp_dual} in more explicit forms, which can readily be implemented and solved on a computer.

\subsection{Bipartite case}

Using the characterisation of Eq.~\eqref{eq:Wsep_cone_bi_v2}, and noting that for $W \in \mathcal L_V$, $W + r \id^\circ$ is also automatically in $\mathcal L_V$, one can write explicitly the primal SDP problem~\eqref{eq:sdp_primal} in the bipartite case as
\begin{equation}\label{eq:sdp_primal_explicit_bi}
\begin{gathered}
 \min \ r \\[1mm]
 \text{s.t.} \quad W + r \, \id^\circ = W^{A \prec B} + W^{B \prec A} \, , \\
 W^{A \prec B} \ge 0 \, , \ \ {}_{[1-B_O]}W^{A \prec B} = 0 \, , \\
 W^{B \prec A} \ge 0 \, , \ \ {}_{[1-A_O]}W^{B \prec A} = 0 \, .
\end{gathered}
\end{equation}

Using now Eq.~\eqref{eq:SV2_characterisation}, the dual SDP problem~\eqref{eq:sdp_dual} writes, more explicitly,
\begin{equation}\label{eq:sdp_dual_explicit_bi}
\begin{gathered}
  \min \ \tr[S \cdot W] \\[1mm]
 \text{s.t.} \quad S = L_V(S^P) \, , \ {}_{B_O}S^P \geq 0 \, , \ {}_{A_O}S^P \geq 0 \, , \\
 \tr[S \cdot \id^\circ] = 1 \, ,
\end{gathered}
\end{equation}
with $L_V$ defined in Eq.~\eqref{def:LV2}.

\subsection{Tripartite case with $d_{C_O} = 1$}

Using Eq.~\eqref{eq:Wsep_cone_tri_v2}, the primal SDP problem~\eqref{eq:sdp_primal} can be written explicitly, in the tripartite case with $d_{C_O} = 1$, as
\begin{equation}\label{eq:sdp_primal_explicit_tri}
\begin{gathered}
 \min \ r \\[1mm]
 \text{s.t.} \quad W + r \, \id^\circ = W^{A \prec B \prec C} + W^{B \prec A \prec C} \, , \\
 W^{A \prec B \prec C} \ge 0 \, , \ \ {}_{[1-B_O]C_I}W^{A \prec B \prec C} = 0 \, , \\
 W^{B \prec A \prec C} \ge 0 \, , \ \ {}_{[1-A_O]C_I}W^{B \prec A \prec C} = 0 \, .
\end{gathered}
\end{equation}

Using now Eqs.~\eqref{eq:S_tri_characterisation} and~\eqref{eq:SV3_characterisation}, the dual SDP problem~\eqref{eq:sdp_dual} writes, more explicitly,
\begin{equation}\label{eq:sdp_dual_explicit_tri}
\begin{gathered}
  \min \ \tr[S \cdot W] \\[1mm]
 \text{s.t.} \quad S = S_{ABC}^P + S_{ABC}^\perp = S_{BAC}^P + S_{BAC}^\perp \, , \\
 S_{ABC}^P \geq 0 \, , \ L_{A \prec B \prec C}(S_{ABC}^\perp) = 0 \, , \\
 S_{BAC}^P \geq 0 \, , \ L_{B \prec A \prec C}(S_{ABC}^\perp) = 0 \, , \\
 S = L_V(S) \, , \ \tr[S \cdot \id^\circ] = 1 \, ,
\end{gathered}
\end{equation}
with $L_{A \prec B \prec C}$, $L_{B \prec A \prec C}$ and $L_V$ defined in Eqs.~\eqref{def:LABC}, \eqref{def:LBAC} and~\eqref{def:LV3}.

\section{Witnesses for the quantum switch}

In this Appendix we give explicit witnesses of the causal nonseparability of the quantum switch.
Although the results reported in the main text do not depend on the initial state $\ket{\psi}$ of the target qubit, the specific form of the witnesses does; in the following we fix it to be $\ket{\psi} = \ket{0}$.

For ease of notations, we will provide the various witnesses in the general form
\begin{equation}
S \, = \, \frac14 \Big( \id + \sum_i \, s_i \, S_i \Big) \,, \label{eq:generic_S}
\end{equation}
for some terms $S_i$ and coefficients $s_i$ to be specified below.
To verify that $S$ is a valid witness, we will provide the explicit decomposition of $S = S_{ABC}^P + S_{ABC}^\perp$ as in~\eqref{eq:S_tri_characterisation} in the form
\begin{equation}
S_{ABC}^\perp \, = \, \frac14 \sum_j \, t_j \, T_j \, , \quad S_{ABC}^P \, = \, S - S_{ABC}^\perp \, , \label{eq:generic_SABCperp}
\end{equation}
for some terms $T_i$ and coefficients $t_i$ to be specified as well. This will allow the reader to check that $L_{A \prec B \prec C}(S_{ABC}^\perp) = 0$ and $S_{ABC}^P \geq 0$, as required by~\eqref{eq:S_tri_characterisation}.

Due to the symmetries of the quantum switch and its witnesses, the second decomposition $S = S_{BAC}^P + S_{BAC}^\perp$ in~\eqref{eq:S_tri_characterisation} can then be obtained as
\begin{equation}
S_{BAC}^\perp \, = \, \mathcal F_{\! A \leftrightarrow B} (S_{ABC}^\perp) \, , \quad S_{BAC}^P \, = \, \mathcal F_{\! A \leftrightarrow B} (S_{ABC}^P) \, ,
\end{equation}
where $\mathcal F_{\! A \leftrightarrow B}$ is the map that exchanges the roles of Alice and Bob, defined as
\begin{eqnarray}
\mathcal F_{\! A \leftrightarrow B}(\sigma_1^{A_I}\otimes\sigma_2^{A_O}\otimes\sigma_3^{B_I}\otimes\sigma_4^{B_O}\otimes\sigma_5^{C_I}) \quad \nonumber \\
= \, \pm \, \sigma_3^{A_I}\otimes\sigma_4^{A_O}\otimes\sigma_1^{B_I}\otimes\sigma_2^{B_O}\otimes\sigma_5^{C_I}
\end{eqnarray}
for $\sigma_i = \id, X, Y, Z$, and where the sign is $+$ if $\sigma_5 = \id$ or $X$, and $-$ if $\sigma_5 = Y$ or $Z$.
(Note that all processes $W$ and all witnesses $S$ considered for the tripartite case in this paper have the symmetry $W = \mathcal F_{\! A \leftrightarrow B} (W)$, $S = \mathcal F_{\! A \leftrightarrow B} (S)$.)

\subsection{Optimal witness with respect to white noise}
\label{app:S_switch}

By solving the dual SDP problem~\eqref{eq:sdp_dual_explicit_tri} for $W = W_\text{switch}$ with CVX, we obtained numerically the witness $S_\text{switch}$ of the form~\eqref{eq:generic_S}, with
{\allowdisplaybreaks
\begin{eqnarray}
& S_1 = Z \id Z \id \id \, , \quad
  S_2 = Z \id \id \id \id + \id \id Z \id \id \, , \nonumber \\[1mm]
& S_3 = \id Z Z \id \id + Z \id \id Z \id \, , \quad
  S_4 = Z Z Z \id \id + Z \id Z Z \id \, , \nonumber \\[1mm]
& S_5 = Z Z \id \id Z - \id \id Z Z Z \, , \quad
  S_6 = \id Z Z \id Z - Z \id \id Z Z \, , \nonumber \\[1mm]
& S_7 = Z \id \id \id Z - \id \id Z \id Z + Z Z \id Z Z - \id Z Z Z Z \, , \nonumber \\
& S_8 = \id Z \id \id Z - \id \id \id Z Z + Z Z Z \id Z - Z \id Z Z Z \, , \nonumber \\[2mm]
& S_9 = \phantom{+} \id X \id X X + \id Y \id Y X + \id X \id Y Y - \id Y \id X Y \nonumber \\
& \phantom{S_{9} =} + \id X  Z  X X + \id Y  Z  Y X + \id X  Z  Y Y - \id Y  Z  X Y \nonumber \\
& \phantom{S_{9} =} +  Z  X \id X X +  Z  Y \id Y X +  Z  X \id Y Y -  Z  Y \id X Y \nonumber \\
& \phantom{S_{9} =} +  Z  X  Z  X X +  Z  Y  Z  Y X +  Z  X  Z  Y Y -  Z  Y  Z  X Y \, , \nonumber \\[1mm]
& S_{10} = \phantom{+} X \id X \id X - X \id X Z X - X Z X \id X + X Z X Z X \nonumber \\
& \phantom{S_{10} =} + Y \id Y \id X - Y \id Y Z X - Y Z Y \id X + Y Z Y Z X \nonumber \\
& \phantom{S_{10} =} +  X \id Y \id Y - X \id Y Z Y - X Z Y \id Y + X Z Y Z Y \nonumber \\
& \phantom{S_{10} =} - Y \id X \id Y + Y \id X Z Y + Y Z X \id Y - Y Z X Z Y \, , \nonumber \\[1mm]
& S_{11} = \phantom{+} \id X X \id \id - \id Y Y \id \id + Z X X \id \id - Z Y Y \id \id \nonumber \\
& \phantom{S_{11} =} + \id X X \id Z - \id Y Y \id Z + Z X X \id Z - Z Y Y \id Z \nonumber \\
& \phantom{S_{11} =}  - \id X X Z Z + \id Y Y Z Z - Z X X Z Z + Z Y Y Z Z \nonumber \\
& \phantom{S_{11} =} + X \id \id X \id - Y \id \id Y \id + X \id Z X \id - Y \id Z Y \id \nonumber \\
& \phantom{S_{11} =} - X \id \id X Z + Y \id \id Y Z - X \id Z X Z + Y \id Z Y Z \nonumber \\
& \phantom{S_{11} =} + X Z \id X Z - Y Z \id Y Z + X Z Z X Z - Y Z Z Y Z \, , \nonumber \\[1mm]
& S_{12} = \phantom{+} X X \id \id X - X X \id Z X + X X Z \id X - X X Z Z X \nonumber \\
& \phantom{S_{12} =} - Y Y \id \id X + Y Y \id Z X - Y Y Z \id X + Y Y Z Z X \nonumber \\
& \phantom{S_{12} =} - X Y \id \id Y + X Y \id Z Y - X Y Z \id Y + X Y Z Z Y \nonumber \\
& \phantom{S_{12} =} - Y X \id \id Y + Y X \id Z Y - Y X Z \id Y + Y X Z Z Y \nonumber \\
& \phantom{S_{12} =} + \id \id X X X - \id Z X X X + Z \id X X X - Z Z X X X \nonumber \\
& \phantom{S_{12} =} - \id \id Y Y X + \id Z Y Y X - Z \id Y Y X + Z Z Y Y X \nonumber \\
& \phantom{S_{12} =} + \id \id X Y Y - \id Z X Y Y + Z \id X Y Y - Z Z X Y Y \nonumber \\
& \phantom{S_{12} =} + \id \id Y X Y - \id Z Y X Y + Z \id Y X Y - Z Z Y X Y \nonumber \\ \label{eq:def_Si}
\end{eqnarray}
}
\hspace{-5.6mm} (where, here and below, the superscripts denoting the different systems are omitted---we keep the order $A_IA_OB_IB_OC_I$---and the tensor products are implicit),
and with the coefficients
\begin{eqnarray}
\begin{array}{lcrlcrlcr}
s_{1} & \!\!\simeq\!\! & 0.2650\, , & s_{2} & \!\simeq\! & 0.6325\, , & s_{3} & \!\!\simeq\!\! & -0.7641\, , \\
s_{4} & \!\!\simeq\!\! & -0.3966\, , & s_{5} & \!\!\simeq\!\! & 0.1168\, , & s_{6} & \!\!\simeq\!\! & 0.2359\, , \\
s_{7} & \!\!\simeq\!\! & 0.0595\, , & s_{8} & \!\!\simeq\!\! & 0.1764\, , & s_{9} & \!\!\simeq\!\! & -0.3340\, , \\
s_{10} & \!\!\simeq\!\! & -0.1128\, , & s_{11} & \!\!\simeq\!\! & 0.1025\, , & s_{12} & \!\!\simeq\!\! & -0.1941 \,.
\end{array} \nonumber \\
\end{eqnarray}

The operator $S_{ABC}^\perp$ is given here by~\eqref{eq:generic_SABCperp}, with
{\allowdisplaybreaks
\begin{eqnarray}
& T_1 = \id Z \id \id \id \, , \quad
  T_2 = \id \id \id Z \id \, , \nonumber \\
& T_3 = Z Z \id \id \id \, , \quad
  T_4 = \id \id Z Z \id \, , \nonumber \\[1mm]
& T_5 = Z \id \id Z \id \, , \quad
  T_6 = \id Z \id Z \id \, , \quad
  T_7 = Z Z \id Z \id \, , \nonumber \\
& T_8 = Z \id Z Z \id \, , \quad
  T_9 = \id Z Z Z \id \, , \quad
  T_{10} = Z Z Z Z \id \, , \nonumber \\[1mm]
& T_{11} = X \id \id X \id + X \id Z X \id - Y \id \id Y \id - Y \id Z Y \id \, , \nonumber \\
& T_{12} = X Z \id X \id + X Z Z X \id - Y Z \id Y \id - Y Z Z Y \id \, , \nonumber \\
& T_{13} = \id X X Z \id + Z X X Z \id - \id Y Y Z \id - Z Y Y Z \id \, , \nonumber \\ \label{eq:def_Tj}
\end{eqnarray}
}
and the coefficients
\begin{eqnarray}
\begin{array}{lcrlcrlcr}
t_{1} & \!\!\simeq\!\! & 0.5157\, , & t_{2} & \!\!\simeq\!\! & -0.2426\, , & t_{3} & \!\!\simeq\!\! & 0.1482\, , \\
t_{4} & \!\!\simeq\!\! & -0.3021\, , & t_{5} & \!\!\simeq\!\! & -1.3741\, , & t_{6} & \!\!\simeq\!\! & 0.4190\, , \\
t_{7} & \!\!\simeq\!\! & 0.7865\, , & t_{8} & \!\!\simeq\!\! & -1.0662\, , & t_{9} & \!\!\simeq\!\! & 0.4785\, , \\
t_{10} & \!\!\simeq\!\! & 0.8460\, , & t_{11} & \!\!\simeq\!\! & -0.5300\, , & t_{12} & \!\!\simeq\!\! & 0.6325\, , \\
t_{13} & \!\!\simeq\!\! & 0.1025 \,.
\end{array} \nonumber \\[-3mm]
\end{eqnarray}

With the witness $S_\text{switch}$ thus defined, we find $\tr[S_\text{switch} \cdot W_\text{switch}] = - r^*_\text{switch} \simeq -1.576 < 0$, as reported in the main text.
Note that in order to measure the witness $S_\text{switch}$, one can decompose each of its terms in a similar way as we did in Subsection~\ref{subsec:W_etas} for $S_{\eta_1, \eta_2}$ in terms of CP maps, implement them and combine the statistics in the appropriate way.

\subsection{A family of witnesses for $W_\text{switch}^\text{depol}(v)$ and $W_\text{switch}^\text{deph}(v)$}
\label{app:S_v}

\begin{figure}
 \vspace{-3mm}
 \includegraphics[width=.9\columnwidth]{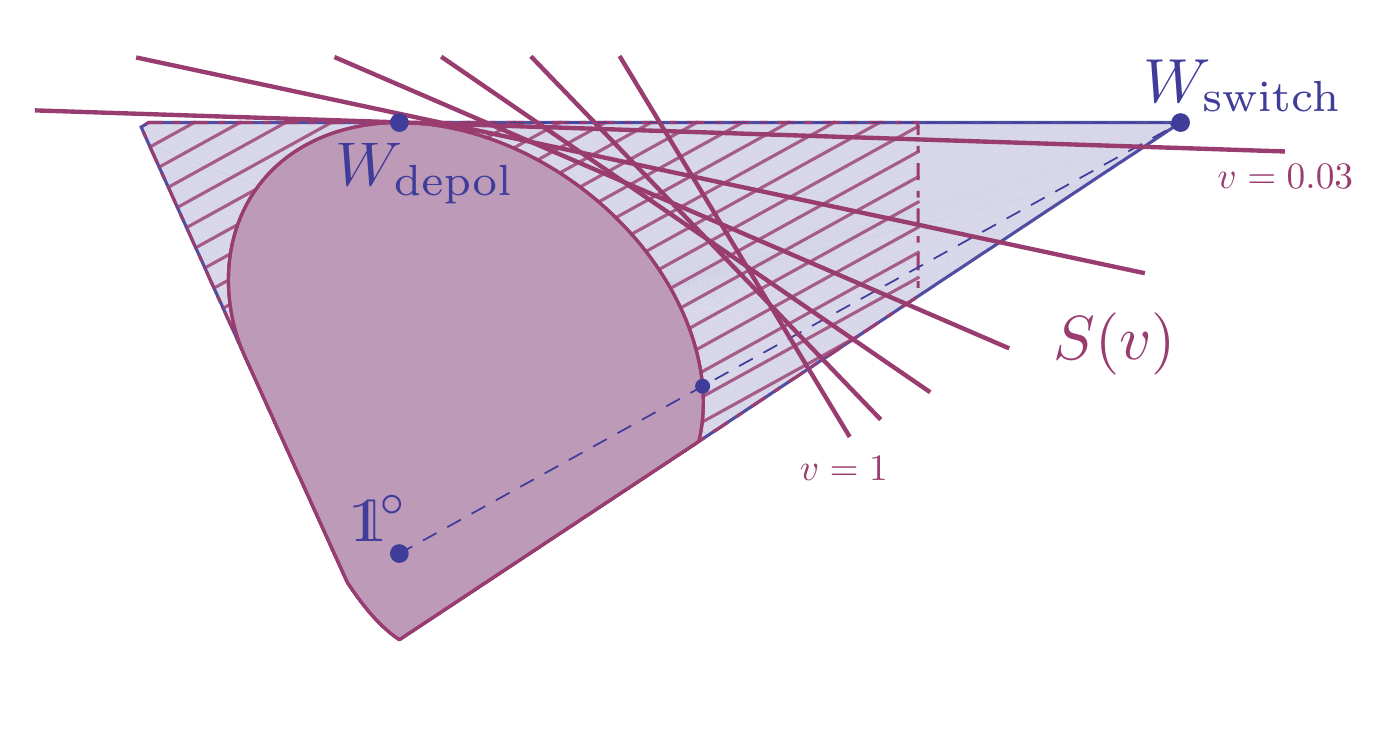}
 \vspace{-5mm}
\caption{Two-dimensional slice of the space of process matrices containing $W_\text{switch}, W_\text{depol}$ and $\id^\circ$, as in Fig.~\ref{fig:W_depol}. Because the line segment containing the processes $W_\text{switch}^\text{depol}(v) = v \, W_\text{switch} + (1{-}v) \, W_\text{depol}$ approaches the set of causally separable processes tangentially, there is no single witness that detects their causal nonseparability for all $v\!>\!0$. Instead, one can use the family of witnesses $S(v)$ of Eq.~\eqref{eq:Sp_explicit}, shown here for the different values of $v = 0.03, 0.2, 0.4, 0.6, 0.8$ and $1$.}
\label{fig:W_depol2}
\end{figure}

\begin{figure}
 \vspace{-3mm}
 \includegraphics[width=.9\columnwidth]{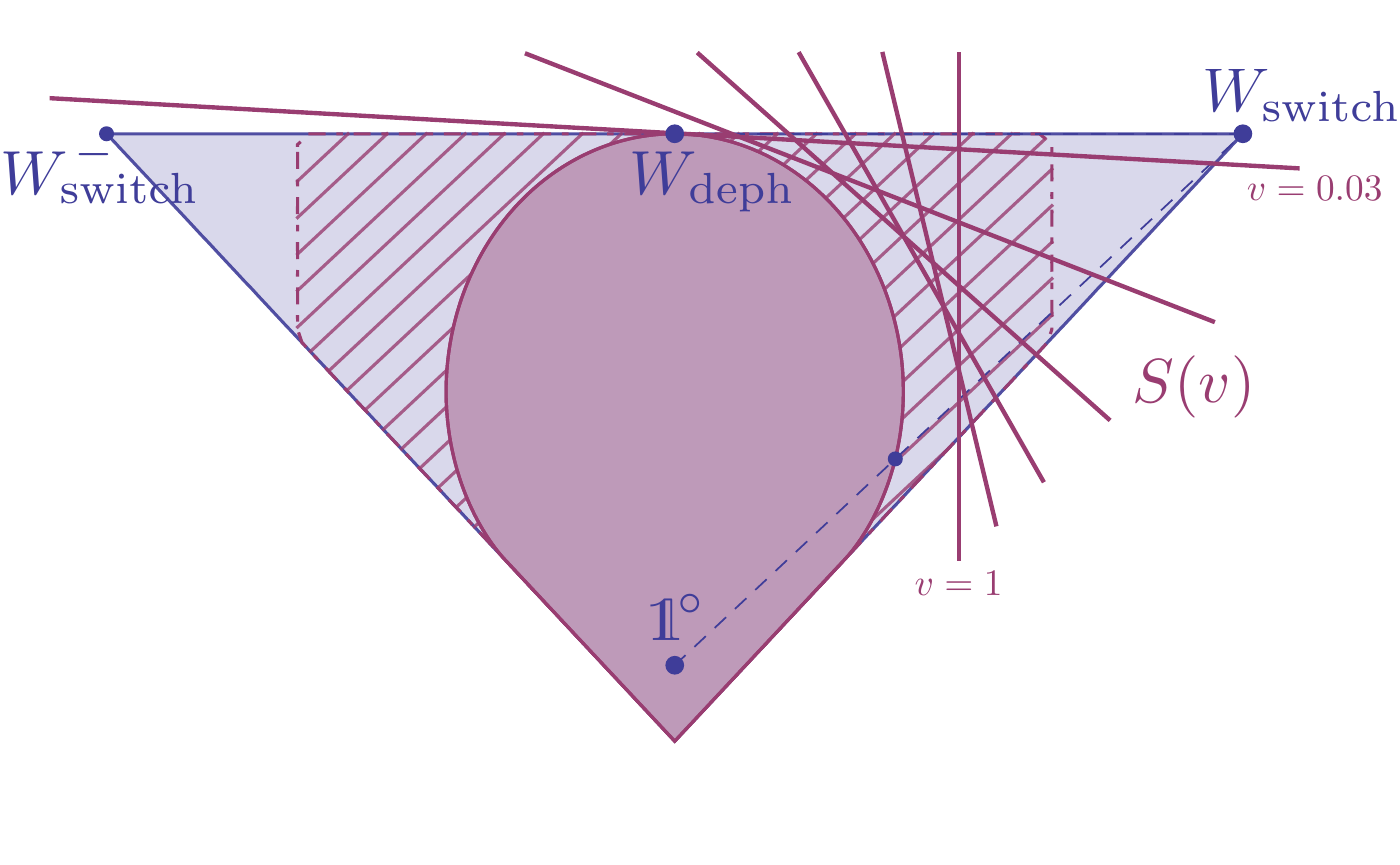}
 \vspace{-7mm}
\caption{Two-dimensional slice of the space of process ma{-}trices containing $W_\text{switch}, W_\text{deph}$ and $\id^\circ$, as in Fig.~\ref{fig:W_deph}. We show here the witnesses $S(v)$ for $v = 0.03, 0.2, 0.4, 0.6, 0.8$ and $1$.}
\label{fig:W_deph2}
\end{figure}

Due to the geometry of the problem, with the line segments containing the processes $W_\text{switch}^\text{depol}(v)$ and $W_\text{switch}^\text{deph}(v)$ being tangent to the set of causally separable processes (see Figures~\ref{fig:W_depol}--\ref{fig:W_deph} or~\ref{fig:W_depol2}--\ref{fig:W_deph2}), one cannot provide a unique witness that would detect the causal nonseparability of all $W_\text{switch}^\text{depol}(v)$ or $W_\text{switch}^\text{deph}(v)$ for all $v > 0$.

Instead, we provide here a family of witnesses $S(v)$, parametrised by $v$. Namely, $S(v)$ and the corresponding $S_{ABC}^\perp(v)$ are given in the forms~\eqref{eq:generic_S} and~\eqref{eq:generic_SABCperp}, with the terms $S_i$ and $T_j$ defined again as in Eqs.~\eqref{eq:def_Si} and~\eqref{eq:def_Tj}, now with the coefficients
\begin{eqnarray}
& s_{1} = s_{2} = 1\, , \quad s_{3} = s_{4} = -\big(1-\frac{v^2}{4}\big)\, , \nonumber \\
& s_{5} = s_{6} = s_{8} = \frac{v^2}{4}\, , \quad s_{9} = -\frac{v}{2} \, ,  \\
& s_{7} = s_{10} = s_{11} = s_{12} = 0\, , \nonumber
\end{eqnarray}
and
\begin{eqnarray}
& t_{6} = t_{7} = t_{9} = t_{10} = 1\, , \nonumber \\
& t_{1} = - t_{2} = t_{3} = - t_{4} = - t_{5}/2 = - t_{8}/2 = 1-\frac{v^2}{4} \, , \nonumber \\
& t_{11} = t_{12} = t_{13} = 0\, . \nonumber \\[-3mm]
\end{eqnarray}
More explicitly, this gives (when written in the order $A_IB_IA_OB_OC_I$ for ease of notation)
\begin{eqnarray}
S(v) & = & \ketbra{0}{0}^{A_I} \ketbra{0}{0}^{B_I} \Big[ \id^{A_OB_OC_I} + \frac{v^2}{4}(Z\id{-}\id Z)^{A_OB_O} Z^{C_I} \nonumber \\
&& \hspace{2.8cm} - \frac{v}{2}(XX+YY)^{A_OB_O} X^{C_I} \nonumber \\
&& \hspace{2.8cm} - \frac{v}{2}(XY-YX)^{A_OB_O} Y^{C_I} \Big] \nonumber \\[-1mm]
&& - \ \frac{1}{2}(1-\frac{v^2}{4}) \Big[ \ketbra{0}{0}^{A_I} Z^{B_I} Z^{A_O} \id^{B_O} \id^{C_I} \nonumber \\[-1mm]
&& \hspace{2.3cm} + Z^{A_I} \ketbra{0}{0}^{B_I} \id^{A_O} Z^{B_O} \id^{C_I} \Big] \, . \label{eq:Sp_explicit}
\end{eqnarray}

One finds
\begin{eqnarray}
\tr[S(v) \cdot W_\text{switch}^\text{depol}(v)] & = & -\big({\textstyle \frac{3-v}{2}}\big)v^2 \, , \\
\tr[S(v) \cdot W_\text{switch}^\text{deph}(v)] & = & -v^2 \, ,
\end{eqnarray}
which give negative values---thus proving that $W_\text{switch}^\text{depol}(v)$ and $W_\text{switch}^\text{deph}(v)$ are causally nonseparable---for all $v > 0$.

Figures~\ref{fig:W_depol2} and~\ref{fig:W_deph2} represent the witnesses $S(v)$, for various values of $v$, in the two-dimensional slices of the space of process matrices containing $W_\text{switch}, W_\text{depol}, \id^\circ$ and $W_\text{switch}, W_\text{deph}, \id^\circ$, respectively.
Note that the witnesses $S(v)$ are not optimal to detect causal nonseparability, as they are not tangent to the set of causally separable processes. E.g., for $v=1$, we find $\tr[S(1) \cdot W_\text{switch}] = -1$, allowing one to prove causal nonseparability of the noisy quantum switch $W_\text{switch}^{\id^{\!\circ}}(v)$~\eqref{eq:qswitch_W_white} only down to $v > 1/2$ (to be compared to $v^*_\text{switch} \simeq 0.3882$ for the optimal witness).
We could not find an analytical expression for optimal witnesses; nevertheless, the witnesses are good enough for our goal, which was to prove that $W_\text{switch}^\text{depol}(v)$ and $W_\text{switch}^\text{deph}(v)$ are causally nonseparable for all $v > 0$.

\subsection{Restricting Alice and Bob's operations \newline to unitaries}
\label{app:S_U}

Here we show how to impose that Alice and Bob's operations are restricted to unitaries, and provide the witness thus obtained.

\subsubsection{Constraints on the CJ representation of a unitary}

Following the convention of~\cite{oreshkov12}, the Choi-Jamio{\l}kowski representation of a unitary operation $U : \mathcal H^{X_I} \to \mathcal H^{X_O}$ is defined as
\begin{equation}
M_U^{X_IX_O} = \big[(\id \otimes U) \, \KetBra{\id} \, (\id \otimes U^\dagger) \big]^T \,,
\end{equation}
where $\id$ is the identity operator on $\mathcal H^{X_I}$, $\Ket{\id} \equiv \Ket{\id}^{X_IX_I} = \sum_{j} \ket{j}^{X_I}\otimes\ket{j}^{X_I} \in \mathcal H^{X_I} \otimes \mathcal H^{X_I}$ is a (nonnormalised) maximally entangled state, $\{\ket{j}^{X_I}\}$ is an orthonormal basis of $\mathcal H^{X_I}$, and $T$ denotes matrix transposition in that basis.

Note first that $U$ is a completely positive and trace-preserving map; the condition $\tr_{X_O} M_U^{X_IX_O} = \id^{X_I}$ that its CJ matrix satisfies, cf Eq.~\eqref{instrument}, can be written as
\begin{equation}
{}_{X_O} M_U^{X_IX_O} = {}_{X_IX_O} M_U^{X_IX_O} \,. \label{eq:contr_MU_X0}
\end{equation}
Let us furthermore calculate:
\begin{eqnarray}
\tr_{X_I} (M_U^{X_IX_O})^T &=& \sum_{i,j,k} \ \bra{i} \otimes U \cdot \ket{j,j} \bra{k,k} \cdot \ket{i} \otimes U^\dagger \nonumber \\
 &=& \sum_{i} \, U \ketbra{i}{i} U^\dagger = \id^{X_O} \,,
\end{eqnarray}
from which it also follows that
\begin{equation}
{}_{X_I} M_U^{X_IX_O} = {}_{X_IX_O} M_U^{X_IX_O} \,. \label{eq:contr_MU_XI}
\end{equation}

\subsubsection{An explicit witness made of unitaries for Alice and Bob}

Solving the SDP problem~\eqref{eq:sdp_dual} with CVX---in its more explicit form~\eqref{eq:sdp_dual_explicit_tri}---after replacing the constraint $S \in \mathcal S_V$ (or $S = L_V(S)$ in the more explicit form) by Eq.~\eqref{eq:constr_U_S}, we obtained numerically the witness $\tilde S$ and the corresponding operator $\tilde S_{ABC}^\perp$ of the forms~\eqref{eq:generic_S}--\eqref{eq:generic_SABCperp}, now with
{\allowdisplaybreaks
\begin{eqnarray}
& S_1 = \id \id \id \id X , \ 
  S_2 = Z Z Z Z X , \ 
  S_3 = Z Z \id \id X \!+\! \id \id Z Z X , \nonumber \\[1mm]
& S_4 = X X X X X \!+\! Y Y Y Y X , \ 
  S_5 = X X Y Y X \!+\! Y Y X X X , \nonumber \\
& S_6 = X Y X Y X \!+\! Y X Y X X , \ 
  S_7 = X Y Y X X \!+\! Y X X Y X , \nonumber \\[1mm]
& S_8 = Z X Z X X \!+\! Z Y Z Y X , \ 
  S_9 = Z X Z Y Y \!-\! Z Y Z X Y , \nonumber \\
& S_{10} = X Z X Z X \!+\! Y Z Y Z X , \ 
  S_{11} = X Z Y Z Y \!-\! Y Z X Z Y , \nonumber \\[1mm]
& S_{12} = X X \id \id X - Y Y \id \id X + \id \id X X X - \id \id Y Y X \, , \nonumber \\
& S_{13} = X Y \id \id Y + Y X \id \id Y - \id \id X Y Y - \id \id Y X Y \, , \nonumber \\[1mm]
& S_{14} = X X Z Z X - Y Y Z Z X + Z Z X X X - Z Z Y Y X \, , \nonumber \\
& S_{15} = X Y Z Z Y + Y X Z Z Y - Z Z X Y Y - Z Z Y X Y \, , \nonumber \\[1mm]
& S_{16} = X Z Z X X - Y Z Z Y X + Z X X Z X - Z Y Y Z X \, , \nonumber \\
& S_{17} = Z X Y Z Y + Z Y X Z Y - X Z Z Y Y - Y Z Z X Y \, , \nonumber \\[1mm]
& S_{18} = X X X Y Y + Y X Y Y Y - X Y X X Y - Y Y Y X Y \, , \nonumber \\
& S_{19} = X X Y X Y + X Y Y Y Y - Y X X X Y - Y Y X Y Y \, , \nonumber \\
\end{eqnarray}
\begin{eqnarray}
& T_1 = Z Z \id \id \id \, , \quad
  T_2 = \id \id Z Z \id \, , \nonumber \\[1mm]
& T_3 = Z \id \id Z \id \, , \quad
  T_4 = \id Z \id Z \id \, , \quad
  T_5 = Z Z Z Z \id \, , \nonumber \\[1mm]
& T_6 = X \id \id X \id - Y \id \id Y \id \, , \ 
  T_7 = X X \id \id \id - Y Y \id \id \id \, , \nonumber \\
& T_8 = \id \id X X \id - \id \id Y Y \id \, , \ 
  T_9 = \id X \id X \id + \id Y \id Y \id \, , \nonumber \\[1mm]
& T_{10} = X X Z Z \id - Y Y Z Z \id \, , \ 
  T_{11} = Z Z X X \id - Z Z Y Y \id \, , \nonumber \\
& T_{12} = Z X Z X \id + Z Y Z Y \id \, , \ 
  T_{13} = Z X X Z \id - Z Y Y Z \id \, , \nonumber \\
& T_{14} = X Z Z X \id - Y Z Z Y \id \, , \ 
  T_{15} = X Z X Z \id + Y Z Y Z \id \, , \nonumber \\[1mm]
& T_{16} = X X X X \id \!+\! Y Y Y Y \id \, , \ 
  T_{17} = X X Y Y \id \!+\! Y Y X X \id \, , \nonumber \\
& T_{18} = X Y X Y \id \!+\! Y X Y X \id , \ 
  T_{19} = X Y Y X \id \!+\! Y X X Y \id , \nonumber \\
\end{eqnarray}
}
and with the coefficients
\begin{eqnarray}
\begin{array}{lcrlcrlcr}
s_{1} & \!\!\simeq\!\! & -0.1396\, , & s_{2} & \!\!\simeq\!\! & -0.2295\, , & s_{3} & \!\!\simeq\!\! & -0.1846\, , \\
s_{4} & \!\!\simeq\!\! & -0.1137\, , & s_{5} & \!\!\simeq\!\! & -0.1262\, , & s_{6} & \!\!\simeq\!\! & -0.2611\, , \\
s_{7} & \!\!\simeq\!\! & -0.0212\, , & s_{8} & \!\!\simeq\!\! & -0.3057\, , & s_{9} & \!\!\simeq\!\! & -0.2157\, , \\
s_{10} & \!\!\simeq\!\! & -0.1044\, , & s_{11} & \!\!\simeq\!\! & -0.0815\, , & s_{12} & \!\!\simeq\!\! & -0.1015\, , \\
s_{13} & \!\!\simeq\!\! & 0.0297\, , & s_{14} & \!\!\simeq\!\! & 0.0979\, , & s_{15} & \!\!\simeq\!\! & -0.1391\, , \\
s_{16} & \!\!\simeq\!\! & -0.0610\, , & s_{17} & \!\!\simeq\!\! & -0.1266\, , & s_{18} & \!\!\simeq\!\! & 0.1150\, , \\
s_{19} & \!\!\simeq\!\! & -0.0570 \,,
\end{array} \nonumber \\[-4mm]
\end{eqnarray}
and
\begin{eqnarray}
\begin{array}{lcrlcrlcr}
t_{1} & \!\!\simeq\!\! & 0.1062\, , & t_{2} & \!\!\simeq\!\! & -0.1387\, , & t_{3} & \!\!\simeq\!\! & -0.4969\, , \\
t_{4} & \!\!\simeq\!\! & 0.4541\, , & t_{5} & \!\!\simeq\!\! & 0.0165\, , & t_{6} & \!\!\simeq\!\! & -0.5239\, , \\
t_{7} & \!\!\simeq\!\! & 0.0269\, , & t_{8} & \!\!\simeq\!\! & -0.0134\, , & t_{9} & \!\!\simeq\!\! & 0.2886\, , \\
t_{10} & \!\!\simeq\!\! & -0.0950\, , & t_{11} & \!\!\simeq\!\! & 0.0102\, , & t_{12} & \!\!\simeq\!\! & 0.0092\, , \\
t_{13} & \!\!\simeq\!\! & 0.1254\, , & t_{14} & \!\!\simeq\!\! & 0.1128\, , & t_{15} & \!\!\simeq\!\! & -0.0439\, , \\
t_{16} & \!\!\simeq\!\! & 0.0994\, , & t_{17} & \!\!\simeq\!\! & 0.0251\, , & t_{18} & \!\!\simeq\!\! & -0.0680\, , \\
t_{19} & \!\!\simeq\!\! & -0.1924 \,.
\end{array} \nonumber \\[-4mm]
\end{eqnarray}

With the witness $\tilde S$ thus defined, we find $\tr[\tilde S \cdot W_\text{switch}] \simeq -0.5058 < 0$. Noting that $\tr[\tilde S \cdot \id^\circ] = \tr[\tilde S \cdot W_\text{depol}] = \tr[\tilde S \cdot W_\text{deph}] = 1$, we find that $\tilde S$ allows one to detect the causal nonseparability of $W_\text{switch}^{\id^{\!\circ}}(v)$~\eqref{eq:qswitch_W_white}, $W_\text{switch}^\text{depol}(v)$~\eqref{eq:qswitch_W_depol} and $W_\text{switch}^\text{deph}(v)$~\eqref{eq:qswitch_W_deph} down to $v = 1/(1-\tr[\tilde S \cdot W_\text{switch}]) \simeq 0.6641$, as reported in the main text.

\medskip

In order to decompose the witness $\tilde S$ in terms of unitaries for Alice and Bob, one can apply for instance the following decomposition to each of its terms $\sigma_1^{A_I}\otimes\sigma_2^{A_O}$ and $\sigma_3^{B_I}\otimes\sigma_4^{B_O}$:
{\allowdisplaybreaks
\begin{eqnarray}
\id \otimes \id & = & {\textstyle \frac12} ( M_{\id} + M_{X} + M_{Y} + M_{Z} ) \, , \nonumber \\
X \otimes X & = & {\textstyle \frac12} ( M_{\id} + M_{X} - M_{Y} - M_{Z} ) \, , \nonumber \\
Y \otimes Y & = & {\textstyle \frac12} ( - M_{\id} + M_{X} - M_{Y} + M_{Z} ) \, , \nonumber \\
Z \otimes Z & = & {\textstyle \frac12} ( M_{\id} - M_{X} - M_{Y} + M_{Z} ) \, , \nonumber \\
X \otimes Y & = & {\textstyle \frac12} ( - M_{P} - M_{PX} + M_{PY} + M_{PZ} ) \, , \nonumber \\
Y \otimes X & = & {\textstyle \frac12} ( - M_{P} + M_{PX} - M_{PY} + M_{PZ} ) \, , \nonumber \\
X \otimes Z & = & {\textstyle \frac12} ( M_{H} + M_{HX} - M_{HY} - M_{HZ} ) \, , \nonumber \\
Z \otimes X & = & {\textstyle \frac12} ( M_{H} - M_{HX} - M_{HY} + M_{HZ} ) \, , \nonumber \\
Y \otimes Z & = & {\textstyle \frac12} ( - M_{HP} + M_{HPX} - M_{HPY} + M_{HPZ} ) \, , \nonumber \\
Z \otimes Y & = & {\textstyle \frac12} ( - M_{PH} + M_{PHX} + M_{PHY} - M_{PHZ} ) \, , \nonumber \\ \\[-5mm] \nonumber
\end{eqnarray}
}
\hspace{-2mm} with $P = \Big(\begin{array}{cc} 1 & 0 \\ 0 & i \end{array}\Big)$ (a phase gate) and $H = \frac{1}{\sqrt{2}}\Big(\begin{array}{cc} 1 & 1 \\ 1 & -1 \end{array}\Big)$ (a Hadamard gate), and where $M_U$ denotes the CJ matrix of the unitary $U$.
(Note that because of Eq.~\eqref{eq:constr_U_S}, no term of the form $\sigma^{A_I} \otimes \id^{A_O}$, $\id^{A_I} \otimes \sigma^{A_O}$, $\sigma^{B_I} \otimes \id^{B_O}$ or $\id^{B_I} \otimes \sigma^{B_O}$ with $\sigma = X, Y$ or $Z$ appears in the decomposition of $\tilde S$.)
Once again, let us emphasise that such decompositions are not unique; one may choose to use a different set of unitaries to decompose $\tilde S$---e.g. one may want to minimise the number of different unitaries to implement (given the dimensions in play, one can do with $10$ for Alice and $10$ for Bob), or the total number of different terms in the decomposition of $\tilde S$.


\end{document}